\documentclass[11.5pt]{article}       

\usepackage {amsfonts,amscd} 

\begin{document}

\sloppy

%

\title{A First Class Constraint Generates Not a Gauge Transformation, But a Bad Physical
Change: The Case of Electromagnetism\footnote{Forthcoming in \emph{Annals of Physics}} }
 \author{J. Brian Pitts %
 \\ University of Cambridge \\jbp25@cam.ac.uk }

\date{20 August 2014} 

\maketitle

\abstract{  In Dirac-Bergmann constrained dynamics, a first-class constraint typically does not \emph{alone} generate a gauge transformation.   By direct calculation it is found that each first-class constraint in Maxwell's theory generates a change in the electric field $\vec{E} $ by an arbitrary gradient, spoiling Gauss's law. The secondary first-class constraint $p^i,_i=0$ still holds, but being a function of derivatives of momenta (mere auxiliary fields), it is not directly about the observable electric field (a function of derivatives of $A_{\mu}),$ which couples to charge.  Only a special combination of the two first-class constraints, the Anderson-Bergmann-Castellani gauge generator $G,$ leaves $\vec{E}$ unchanged.  Likewise only that combination leaves the canonical action invariant---an argument independent of observables.  
If  one uses a first-class constraint to generate instead a {canonical transformation}, one partly strips the canonical coordinates of physical meaning as electromagnetic potentials, vindicating the Anderson-Bergmann  Lagrangian orientation of interesting canonical transformations.
 The need to keep gauge-invariant the relation $\dot{q}- \frac{\delta H}{\delta p}= -E_i -p^i=0$   supports using the gauge generator and  primary Hamiltonian rather than the separate first-class constraints and the extended Hamiltonian. 
 
Partly paralleling Pons's criticism,  it is shown that 
Dirac's proof that a first-class primary constraint generates a gauge transformation, by comparing evolutions from \emph{identical} initial data, cancels out and hence fails to detect the alterations made to the initial state.  It also neglects the arbitrary coordinates multiplying the secondary constraints \emph{inside} the canonical Hamiltonian.
Thus the gauge-generating property has been ascribed to the primaries alone, not the primary-secondary team  $G$. 
Hence the Dirac conjecture about secondary first-class constraints as generating gauge transformations rests upon a false presupposition about primary first-class constraints.   Clarity about Hamiltonian electromagnetism will be useful for an analogous treatment of GR.


 }


\tableofcontents


\section{Introduction}

In the early stages of research into constrained Hamiltonian dynamics by Bergmann's school and earlier by Rosenfeld \cite{RosenfeldQG}, it was important to ensure that the new Hamiltonian formalism agreed with the established Lagrangian formalism.  That was very reasonable, for what other criteria for success were there at that stage?  One specific manifestation of Hamiltonian-Lagrangian equivalence was the recovery of the usual 4-dimensional Lagrangian gauge transformations for Maxwell's electromagnetism and (more laboriously) GR by Anderson and Bergmann \cite{AndersonBergmann}.  4-dimensional Lagrangian-equivalent gauge transformations were implemented by Anderson and Bergmann in the Hamiltonian formalism using the gauge generator (which I will call $G$), a specially tuned sum of the first-class constraints, primary and secondary, in electromagnetism or GR \cite{AndersonBergmann}.    

At some point, %
equivalence with 4-dimensional Lagrangian considerations came to play a less significant role.  Instead the idea that a first-class constraint \emph{by itself} generates a gauge transformation became increasingly prominent. That claim \cite{DiracLQM}
 has been called the ``{`standard'}'' interpretation \cite{GotayIneffective} and is officially adopted in Henneaux and Teitelboim's book \cite[pp. 18, 54]{HenneauxTeitelboim} (at least nominally, though not always in reality \cite{PonsDirac})
and in countless other places \cite{CarlipQGreview,DittrichPartialConstrained,RotheRothe}. This idea displaced the Anderson-Bergmann gauge generator until the 1980s and remains a widely held view, though no longer a completely dominant one in the wake of the Lagrangian-oriented reforms of Castellani, Sugano, Pons, Salisbury, Shepley, \emph{etc.}  Closely paralleling the debate between the Lagrangian-equivalent gauge generator $G$ and the distinctively Hamiltonian idea that a first-class constraint generates a gauge transformation is the debate between the Lagrangian-equivalent primary Hamiltonian $H_p$ (which adds to the canonical Hamiltonian $H_c$ all the primary constraints, whether first- or second-class) and Dirac's extended Hamiltonian $H_E$, which adds to the primary Hamiltonian the first-class secondary constraints.

 A guiding theme of Pons, Shepley, and Salisbury's series of works   \cite{PonsSalisburyShepley,ShepleyPonsSalisburyTurkish,PonsSalisbury} 
is important:
\begin{quote}  
We have been guided by the principle that the Lagrangian and Hamiltonian formalisms should be equivalent   \ldots in coming to the conclusion that they in fact are. \cite{PonsReduce}
\end{quote}

 While proponents of the primary Hamiltonian $H_p$ 
have emphasized the value of making the Hamiltonian formalism equivalent to the Lagrangian, what has perhaps been lacking until now is an effective argument that the Lagrangian-inequivalent extended Hamiltonian is erroneous.  While inequivalence of the extended Hamiltonian to the Lagrangian might seem worrisome, it is widely held that the difference is confined to gauge-dependent unobservable quantities and hence makes no real physical difference.  
  If that claim of empirical equivalence were true, it would be a good defense of the permissibility of extending the Hamiltonian.  But is that claim of empirical equivalence true?

This paper shows that the Lagrangian-equivalent view of the early Anderson-Bergmann work \cite{AndersonBergmann} and the more recent Lagrangian-oriented reforms are correct, that is, mandatory rather than merely an interesting option.  It does so by showing by direct calculation that a first-class constraint makes an observable difference to the observable electrical field, indeed a bad difference:  it spoils Gauss's law $\nabla \cdot \vec{E}=0.$  The calculation is perhaps too easy to have seemed worth checking.  Analogous calculations with the canonical action give a similar verdict.  

 This paper also critiques the usual Hamiltonian-focused views of observables deployed in the extended Hamiltonian tradition to divert attention from such a calculation or (in the case of one paper that calculates the relevant Poisson brackets \cite{CostaDiracConjecture}) to explain away the embarrassment of a Gauss's law-violating change in the electric field.  Attention is paid to which variables have physical meaning when (off-shell \emph{vs.} on-shell), \emph{etc.}, with the consequence that canonical momenta have observable significance only derivatively and on-shell rather than primordially and off-shell.  The fact that introducing a Hamiltonian formalism neither increases nor decreases one's experimental powers is implemented consistently.  
Indeed apart from constraints, canonical momenta (the ones replacing the solved-for velocities, such as $\dot{A}_i$, $i=1,2,3$ for electromagnetism) play the role of auxiliary fields in the Hamiltonian action $\int dt(p\dot{q}-H(q,p))$:  one can vary with respect to $p$, get an equation $\dot{q}-\frac{\delta H}{\delta p}=0$ to solve for $p$, and then use it to eliminate $p$ from the action, getting $\int dt L$ \cite{HenneauxAuxiliary}.
One would scarcely call an auxiliary field a primordial observable and the remaining dependence on $q$ or its derivatives in $L$ derived.

This paper interacts with the argument that a first-class primary constraint generates a gauge transformation.  This argument has been copied in various places, including several more recent books \cite{Govaerts,HenneauxTeitelboim,RotheRothe}.  Pons has critiqued this derivation authoritatively \cite{PonsDirac}; my critique offers a  complementary perspective on the logic of Dirac's argument.
 One can see by inspection that the 3-vector potential $A_i$ is left alone by the sum of first-class primary constraints, while the scalar potential is changed.  But the science of electrostatics \cite{Jackson} explores the physical differences associated with different scalar potentials $A_0$ and the same (vanishing) 3-vector potential $A_i.$ Thus Dirac has pronounced observably different electric fields to be gauge-related.  The  mistake can be seen as involving a failure to attend to the term $-A_0 p^i,_i$ in the canonical Hamiltonian density for electromagnetism (to apply Dirac's general analysis to that specific case) in some cases where it cannot be ignored.  Apparently thinking that the secondary constraints were absent or cancelled out in different evolutions (which they do not in general because the coefficient $-A_0$ of the secondary constraint is gauge-dependent), Dirac felt the need to add in the secondary first-class constraints by hand, extending the Hamiltonian, in order to recover the gauge freedom that supposedly was missing.  Thus the motivation for the extended Hamiltonian and the original `proof' that primary first-class constraints generate gauge transformations are undermined.

This paper also explores the consequences for Dirac's conjecture that all first-class secondary constraints generate gauge transformations.  That conjecture was predicated on the assumed validity of the proof that primary first-class constraints generate gauge transformations.  Hence the Dirac conjecture cannot get started; it rests on a false presupposition.


The most obvious interesting examples of first-class constraints, as in Maxwell's electromagnetism and in General Relativity, \emph{change the physical state or history}, and in a bad way, spoiling the Lagrangian constraints, the constraints in terms of $q$ and $\dot{q}$.  Those are the physically relevant constraints, parts of Maxwell's equations (Gauss's law) or the Einstein equations; the (unconstrained) canonical momenta $p$ are merely auxiliary quantities useful insofar as they lead back to the proper behavior for $q$ and $\dot{q}.$  
While there are examples where a first-class constraint does generate a gauge transformation,\footnote{A free relativistic particle with all 4 coordinates as dynamical functions of an arbitrary parameter, but without an auxiliary lapse function $N$, is an example kindly mentioned by Josep Pons.  If one has the auxiliary lapse function \cite{PonsSalisburyShepley,PonsSalisburySundermeyerFolklore}, one gets a primary and a secondary constraint, the latter including a piece quadratic in momenta---looking naively like a Hamiltonian, one might say.  If one instead integrates out the lapse using $\frac{\partial L}{\partial N}=0$, then the resulting Hamiltonian formalism has vanishing canonical Hamiltonian, while the primary constraint becomes more interesting. %
  Conserving the primary constraint  gives no secondary or higher constraint, partly because the canonical Hamiltonian vanishes. The solitary primary constraint is first-class by antisymmetry of the Poisson bracket.  In the absence of higher-order constraints, the gauge generator is just the smeared primary first-class constraint, so in this case a primary constraint does indeed generate a gauge transformation.   
A free relativistic particle is of course a system for which nothing happens.  Potentially more interesting is the fact that one can integrate out the lapse in GR
as in the Baierlein-Sharp-Wheeler action. Then the Hamiltonian constraint arises at the primary level \cite{BarbourFOM}.} such cases are  rare or uninteresting in comparison to those that do not. 
Instead, a gauge transformation is generated by a \emph{special combination} of first-class constraints, namely, the gauge generator $G$ \cite{AndersonBergmann,CastellaniGaugeGenerator,SuganoGaugeGenerator,PonsSalisburyShepleyYang}. 
It long was easy to neglect 4-dimensional coordinate transformations in GR because a usable gauge generator was unavailable after the $3+1$ split innovation in 1958 \cite{AndersonPrimary,DiracHamGR} rendered the original (rather fearsome) $G$  \cite{AndersonBergmann} obsolete by trivializing the primary constraints.  The $3+1$ gauge generator $G$ finally appeared in  1982 \cite{CastellaniGaugeGenerator}, the lengthy delay suggesting that no one was looking for it for a long time.

For Maxwell's electromagnetism, where everyone knows what a gauge transformation is---what makes no physical difference, namely, leaving  $\vec{E}$ and $\vec{B}$ unchanged---and where all the calculations are easy, one can \emph{test} the claim that a first-class constraint generates a gauge transformation.  There is no room for ``interpretation,'' ``definition,'' ``assumption,'' ``demand,'' or the like.  Additional postulates are either redundant or erroneous.  Surprisingly, given the age of the claim, %
such a test apparently hasn't been made before, at least not completely and successfully, 
and has rarely been attempted.
Perhaps the temptation to default to prior knowledge has been irresistible.  By now the sanction of tradition and authority also operate.  Views about observability have also deflected attention away from the question in the context of the extended Hamiltonian.  
 Anyway the test can be made by re-mathematizing the verbal formula.  The result is clearly negative:  a first-class constraint---either the primary or the secondary---generates a physical difference, a change in $\vec{E}.$  This change involves the gradient of an arbitrary function, implying that $\nabla \cdot \vec{E}\neq 0,$ spoiling Gauss's law.  Similar problems arise in GR, as will be discussed in a subsequent work in preparation.

While the process of Lagrangian-equivalent reform started some time ago, it has by no means swept the field.  One also finds works that inconsistently mix the two views.  
While such issues cause little trouble in electromagnetism because all calculations are easy and one already knows all the right answers anyway and so does not depend on the Hamiltonian formalism, it does matter for GR, where 
the right answers are sometimes unknown or controversial and many calculations are difficult.  
It is therefore important both to show that the extended Hamiltonian formalism and associated view of gauge freedom are incorrect (as this paper does) and to implement consistently the consequences of the Lagrangian-equivalent Hamiltonian formalism in the arenas of change and observables in GR (as successor papers will do).

\section{First-Class Primary Constraint \emph{vs.}  Gauge Transformation}  %

It is widely held that a primary first-class constraint generates a gauge transformation  \cite[p. 21]{DiracLQM} \cite[p. 17]{HenneauxTeitelboim} \cite{WipfBadHonnef} \cite[p. 68]{RotheRothe}. 
In a later section a tempting error that leads to this conclusion, namely, neglecting the fact that first-class secondary constraints with gauge-dependent coefficients already appear in the canonical Hamiltonian, will be discussed.  For now a direct and apparently novel (surprisingly enough) test will be applied to show simply \emph{that} the transformation effected by a first-class primary constraint is not generally a gauge transformation.  The test is simply ascertaining what happens to the electric field in Maxwell's electromagnetism, the standard example of a simple yet physically relevant relativistic field theory.

The electromagnetic field strength  $F_{\mu\nu} =_{df}  \partial_{\mu} A_{\nu} - \partial_{\nu}  A_{\mu}$ is unchanged by $A_{\mu} \rightarrow A_{\mu} - \partial_{\mu} \epsilon$.  
 $\vec{E}$ and $\vec{B}$ are parts of $F_{\mu\nu}$  and hence constructed from derivatives of $A_{\mu}.$ 
(For a charged particle in an electromagnetic field, or for a charged scalar field interacting with the electromagnetic field, it is the derivatives of $A_{\mu}$, not the canonical momentum conjugate to $A_{\mu},$ to which charge responds.) That fact will prove important once, in the Hamiltonian formulation, one has conceptually independent canonical momenta $p^i$ satisfying the secondary first-class constraint $p^i,_i=0.$  Electromagnetic gauge transformations are defined ``off-shell,'' without assuming the field equations---in other words, on kinematically possible trajectories, not just dynamically possible trajectories (to use terminology from (\cite{Anderson}). The field equations in question are $$\dot{A}_i - \frac{ \delta H}{ \delta p^i } = -E_i - p^i =0.$$
 But off-shell there is no relationship between $\dot{A}_i$  and $p^i$, and hence none between $\vec{E}$ and $p^i$. The constraint $p^i,_i=0$ in phase space can cease to be equivalent to the Lagrangian constraint $\nabla \cdot \vec{E} =0$ if one does something inadvisable---such as treating $p^0$ or $p^i,_i$ as if it (by itself) generated a gauge transformation.  That is somewhat as Anderson and Bergmann warned in discussing canonical transformations that do not reflect Lagrangian invariances:  the meanings of the canonical coordinates and/or momenta can be changed \cite{AndersonBergmann}.   The relationship between first-class constraints, the gauge generator $G$, and canonical transformations will be explored below.  It turns out that $G$ does basically the same good thing whether one simply takes Poisson brackets directly or makes a canonical transformation; a first-class constraint does either something permitted but pointless (a position-dependent field redefinition) or something bad (spoiling Gauss's law).

The Legendre  transformation from $\mathcal{L}$ and  $\dot{A}_{\mu}$ to $\mathcal{H}$ and $p^{\mu}$ fails because  $p^{\mu}=_{df} \frac{\partial \mathcal{L} }{\partial \dot{A}_{\mu} } $ is not soluble for $\dot{A}_{\mu}$ \cite{Sundermeyer}.  One gets a primary constraint  
  $p^0(x)=_{df} \frac{\partial \mathcal{L} }{\partial A_0,_0 }=0. $ 
Likewise in General Relativity  \cite{AndersonPrimary,DiracHamGR}, one can choose a divergence in $\mathcal{L}$ and a set of fields using a $3+1$ split,  the lapse $N = 1/\sqrt{-g^{00}}$ and shift vector $N^i={^{3}g}^{ij}g_{j0}$, such that   $p_{0}=_{df} \frac{\partial \mathcal{L} }{\partial {N},_{0} }=0$ and  $p_{i}=_{df} \frac{\partial \mathcal{L} }{\partial {N}^i,_{0} }=0$.
One needs the dynamical preservation of the primary constraints, from which emerge secondary constraints.  In electromagnetism this constraint is Gauss's law, or rather, something equivalent to Gauss's law using $\dot{A}_{i} = \frac{ \delta H}{\delta p^{i} }.$  
 The algorithm of constraint preservation terminates thanks to the constraint algebra.  
 The time evolution is under-specified: there is  gauge/coordinate freedom  due to the presence of first-class constraints (having $0$ Poisson brackets among themselves,  strongly in electromagnetism, using the constraints themselves in GR).  All constraints in both theories are first-class.  
The Poisson bracket is  $$\{ \phi(x), \psi(y) \}=_{df} 
 \int d^3z \sum_A \left( \frac{\delta \phi(x) }{\delta q^A(z) } \frac{\delta \psi(y)}{\delta p_A(z)}  - \frac{\delta \phi(x)}{\delta p_A(z)} \frac{\delta \psi(y)}{\delta q_A(z)} \right);$$ the fundamental ones are $\{ q^A(x), p_B(y) \} = \delta^A_B \delta(x,y).$

These familiar matters set up the belated \emph{test} of whether a first-class constraint really generates a gauge transformation.    \emph{Exactly what} do first-class constraints have to do with gauge freedom?  Curiously, this question has two standard but incompatible answers in the literature on constrained dynamics.   One of them holds  that the gauge generator $G$ \cite{AndersonBergmann,CastellaniGaugeGenerator,SuganoGaugeGenerator,PonsSalisburyShepleyYang} generates a gauge transformation, a change in the description of the physical state (or history, if GR is the theory in question) that makes no objective difference.\footnote{The history, not the state, is the more fundamental choice here.  The idea of a `state' is even ambiguous in that Hamiltonian states use thinner moments (quantities at a single instant) than do Lagrangian states (quantities and their first time derivatives). } 
  This answer is motivated by Hamiltonian-Lagrangian equivalence and is associated with the primary Hamiltonian.  It was eclipsed during the 1950s or 1960s and has slowly reappeared since the 1980s.  
The other standard answer, more influential in the literature on canonical GR, is that a first-class constraint (by itself) generates a gauge transformation,
a distinctively Hamiltonian claim, one that is often  associated with the extended Hamiltonian.

In electromagnetism the fundamental Poisson brackets are $\{ A_{\mu}(x), p^{\nu}(y) \}= \delta^{\nu}_{\mu} \delta(x,y):$ a `big' Poisson bracket that includes $A_0$ and $p^0$.
The constraints are the primary  $p^0(x)=0$  and the secondary  $p^i,_i(x)=0$. One hopes to keep the latter equivalent to Gauss's law, but that isn't automatic because Gauss's law involves the electric field, whereas the secondary constraint involves a canonical momentum, which \emph{a priori} is unrelated to the electric field and becomes equal to it (up to a sign, depending on one's conventions) only using the equations of motion $\dot{q} = \frac{\delta H}{\delta p}.$

 What does $p^0(x)$  do?  By re-mathematizing the claim that a first-class constraint generates a gauge transformation, one predicts from it that $p^0(x)$ changes $A_{\mu}$ \emph{via}  a gauge transformation.   Smearing  $p^0(y)$ with arbitrary  $\xi(t,y)$ and taking the Poisson bracket gives \cite[p. 134]{Sundermeyer} %
\begin{eqnarray}  \delta A_{\mu}(x) = \{ A_{\mu}(x), \int d^3y p^0(y) \xi(t,y) \} = \delta^0_{\mu} \xi(t,x).\end{eqnarray}  
While this expression doesn't look just as one would expect from experience with the Lagrangian, might it reflect (as is often claimed abstractly) some more general gauge invariance disclosed by the Hamiltonian (especially the extended Hamiltonian) formalism?  One can calculate that   \begin{eqnarray} \delta F_{\mu\nu } =_{df} F_{\mu\nu}[A+\delta A] - F_{\mu\nu}[A] =  \partial_{\mu} \delta A_{\nu}  - \partial_{\nu} \delta A_{\mu} =  \partial_{\mu}\xi \delta_{\nu}^0  - \partial_{\nu}\xi \delta_{\mu}^0. \end{eqnarray} 
This definition reflects the standard gauge variation of a velocity as the time derivative of the gauge variation of the corresponding coordinate. %
Letting $\mu=m,$ $\nu = n$ (Latin letters running from $1$ to $3$), one sees that the magnetic field is invariant \cite[p. 134]{Sundermeyer}, which is a good sign. %

 What happens to the electric field $\vec{E}$?  Here Sundermeyer, having come so far, stops short \cite[p. 134]{Sundermeyer}.\footnote{Costa \emph{et al.} \cite{CostaDiracConjecture} got the same mathematical result.  They failed to discern that it was problematic physically, for reasons discussed below involving which fields are observable.} 
 Let $\mu=0,$ $\nu=n$: \begin{eqnarray} \delta F_{0n } = -\delta \vec{E} = \partial_{0} \delta A_{n}  - \partial_{n} \delta A_{0} =  \partial_{0}\xi \delta_{n}^0  - \partial_{n}\xi \delta_{0}^0 =  - \partial_{n}\xi. \label{PFCVary}  \end{eqnarray}   Unless one restricts oneself to the very uninteresting special case of spatially constant $\xi$ (perhaps still depending on time), this is not a gauge transformation, because the world is different, indeed worse.\footnote{This result shows the inadequacy of the idea that a first-class constraint generates a time-independent gauge transformation. Even a time-independent $\xi(x)$ changes $\vec{E}$ and spoils Gauss's law. Things fare somewhat better for that idea for the secondary constraint. }  
While $\vec{B}$ is unchanged, $\vec{E}$ is changed by $ \partial_n \xi(t,x).$ 
 Thus Gauss's law $\nabla \cdot \vec{E} = 0$ is spoiled:  $\nabla \cdot \vec{E} = \nabla^2 \xi \neq 0$ typically. This spoilage of the Lagrangian constraint is not immediately obvious because the secondary constraint $p^i,_i=0$  still holds.  The trouble is that this expression, which lives in phase space, ceases to mean what one expected.  $p$ is   independent of $q,$ but $\dot{q}$ is dependent on $q$ by definition; hence $\dot{q}$ and $p$ are independent, at least until \emph{after} Poisson brackets are calculated.  $\vec{E}$ is a familiar function of derivatives of $A_{\mu};$ the  change in $A_{\mu}$ implies a Gauss's law-violating change in $\vec{E}.$ While  still $p^i,_i=0$ (the phase space
constraint surface is preserved), this constraint is no longer equivalent to Gauss's law:   $p^i,_i=0$ but  $\nabla \cdot \vec{E} \neq 0.$ 
 Instead $\vec{E}$ acts as though some phantom charge density were a source.  The relationship between $p$ and $\dot{q}$ has been altered, something that Anderson and Bergmann warned could happen \cite{AndersonBergmann}.   Changing $\vec{E}$ is  a physical difference, not a gauge transformation---indeed a bad physical difference,  spoiling Gauss's law.

Sundermeyer commented on the ``vague relation between first class constraint transformations and local gauge transformations.''  \cite[p. 134]{Sundermeyer}   He  appeared to be in the process of reinventing the gauge generator in the chapters on electromagnetism and Yang-Mills theories \cite[pp. 134, 168]{Sundermeyer}.
%


Others have fallen into  error on this point.  Bergvelt and de Kerg, 
applying their Hamiltonian technique to a Yang-Mills field, 
\begin{quote}
\ldots first note that two points of [final constraint manifold] $M_2$ of the form $(A_0, A, \pi)$ and $(\hat{A}_0, A, \pi)$ (i.e. differing only in their $A_0$-component) are gauge equivalent.  They can be connected by an integral curve of the gauge vector field $\dot{A}(\frac{\delta}{\delta A_0 })$, with $\dot{A} = \hat{A}_0-A_0$.  So the $A_0$-component of points of $M_2$ is physically irrelevant and without loss of generality we can ignore it.  \cite[p. 133]{BergveltHamYM2}.  \end{quote}
But electrostatics studies what electric fields  can be generated by merely the scalar potential \cite{Griffiths,Jackson}. 
One can of course change $A_0$ freely, but only by paying the price by changing $A_{\mu}$ to compensate.  
 Presumably the ``crucial \emph{assumption}'' that some freedom located in their preceding paper had no physical significance \cite{BergveltHamYM} contributed to this difficulty.  One already knows from the Lagrangian formulation what the gauge freedom is, so there is no room for independent postulates; they are either redundant or erroneous.  Gotay, Nester and Hinds, following Dirac, make a similar mistake with the primary constraint \cite{GotayNesterHinds}, as will appear shortly.  

\section{First-Class Secondary Constraint \emph{vs.} Gauge Transformation}

What does the secondary constraint $p^i,_i(x)$ do?  According to Henneaux and Teitelboim, excepting a few exotic counterexamples, 
\begin{quote}
\emph{one postulates, in general, that all first-class constraints generate gauge transformations}.  This is the point of view adopted throughout this book.  There are a number of good reasons to do this.  First, the distinction between primary and secondary constraints, being based on the Lagrangian, is not a natural one from the Hamiltonian point of view.\ldots Second, the scheme is consistent.\ldots  Third, as we shall see later, the known quantization methods for constrained systems put all first-class constraints on the same footing, \emph{i.e.}, treat all of them as gauge generators.  It is actually not clear if one can at all quantize otherwise.  Anyway, since the conjecture holds in all physical applications known so far, the issue is somewhat academic.  (A proof of the Dirac conjecture under simplifying regularity conditions that are generically fulfilled is given in subsection 3.3.2.) 
\cite[p. 18, emphasis in the original]{HenneauxTeitelboim}
\end{quote}
This is a striking passage.  
 Getting sensible results about observables does require privileging the Lagrangian formalism, so one should not downplay the primary \emph{vs.} secondary distinction on Hamiltonian grounds. Fortunately the constraints wind up being put to work together rather than separately later in the work \cite{PonsDirac}.

One can simply calculate what the secondary constraint $p^i,_i$ does to the electric field.  To my knowledge, this has not been done, surprisingly enough, or at least not done successfully and then appropriately and consistently understood.\footnote{Two cases where the calculation was done are (\cite{CostaDiracConjecture,Deriglazov1}). The former paper will be discussed shortly.  The latter has an ambiguous attitude toward the individual \emph{vs.} team use of first-class constraints.  Without treatment of coupling to charge, the physically measurable quantities in the formalism are not obvious. }  (Proponents of the primary Hamiltonian and its gauge generator don't need to calculate it, because the usual gauge transformation of $A_{\mu}$ to $A_{\mu} - \partial_{\mu} \epsilon$ makes the answer obvious.  Only proponents of the extended Hamiltonian and/or the associated claim that a first-class constraint generates a gauge transformation ought to have done so.  Costa \emph{et al.} did perform relevant calculations on this point \cite{CostaDiracConjecture}; the reason that they did not discern the absurdity of the result involves observables and will be discussed below.) The answer is the secondary first-class constraint  also changes $\vec{E},$ also generally violating Gauss's law, at least if one uses a time-dependent smearing function. If one does not use time-dependent smearing functions, then one has no way to write $G$ and hence no hope of recovering the usual electromagnetic gauge transformations.   Part of the trouble, according to Pons \cite{PonsDirac}, is that Dirac envisaged gauge transformations as pertaining to 3-dimensional hypersurfaces, whereas Bergmann tended to envision them (more appropriately for GR given the freedom to slice more or less arbitrarily) as pertaining to 4-dimensional histories. %
Smearing $p^i,_i$ with an arbitrary function $\epsilon(t,y)$, one finds \cite{CostaDiracConjecture,WipfBadHonnef}
\begin{eqnarray} \delta A_{\mu}(x) = \{ A_{\mu}(x), \int d^3y p^i,_i(y) \epsilon(t,y) \} = -\delta_{\mu}^i \frac{ \partial}{\partial x^i} \epsilon(t,x). \end{eqnarray}
One can thus find the change in $F_{\mu\nu}$:  
 \begin{eqnarray}  \delta F_{\mu\nu } = \partial_{\mu} \delta A_{\nu}  - \partial_{\nu} \delta A_{\mu} =  \partial_{\mu} (-\delta_{\nu}^i \frac{ \partial}{\partial x^i} \epsilon)   - \partial_{\nu} (-\delta_{\mu}^i \frac{ \partial}{\partial x^i} \epsilon) =  \delta_{\mu}^i  \partial_{\nu} \partial_i \epsilon  -   \delta_{\nu}^i  \partial_{\mu} \partial_i \epsilon.  \end{eqnarray}
Clearly $\vec{B}$  is unchanged, but $\vec{E}$'s change is obtained by setting $\mu=0,$ $\nu=n$:
\begin{eqnarray} \delta F_{0n} = -\delta \vec{E}= \delta_{0}^i  \partial_{n} \partial_i \epsilon  -   \delta_{n}^i  \partial_{0} \partial_i \epsilon= -   \partial_n \partial_{0} \epsilon. \end{eqnarray}   Again $\vec{E}$ is changed by an arbitrary gradient, and Gauss's law is spoiled: $\nabla \cdot \vec{E} = \nabla^2 \dot{\epsilon}.$  One could avoid this change in $\vec{E}$ using exclusively time-independent smearing functions; but one will thereby fail to recover the usual electromagnetic gauge transformations.  Imposing time-independence (or spatial homogeneity) on smearing functions is of course also incompatible with Lorentz invariance (to say nothing of general covariance for the analogous issue in GR).

 So neither constraint by itself generates a gauge transformation (without a pointless and misleading restriction on smearing, at any rate, which restricts what the constraint itself is trying to generate).  Each makes a bad physical difference.  Dirac wrote that ``I haven't found any example for which there exists first-class secondary constraints which do generate a change in the physical state.'' \cite[p. 24]{DiracLQM}  
 But  Castellani rightly said that 
\begin{quote}
Dirac's conjecture that all secondary first-class constraints generate
symmetries is revisited and replaced by a theorem.\ldots  The old question whether secondary first-class constraints generate gauge
symmetries or not \ldots is then solved: they are \emph{part} of a gauge generator
$G$ \ldots \cite[pp. 357, 358]{CastellaniGaugeGenerator}. (emphasis in the original)  
\end{quote}
After many years the force of the word ``replaced'' still has not been absorbed:  it involves the \emph{elimination of the old erroneous claim}, not just the introduction of a new true claim. Perhaps Castellani's diplomatic wording has slowed the understanding of his result.   His target was the secondaries in isolation (supposedly the live issue \emph{vis-a-vis} the Dirac conjecture), but the same holds for the primaries.  Neither generates a gauge transformation by itself, but the two together, properly tuned, do.  


One can find examples where these problems should have been noticed.  One is the influential  paper by Gotay, Nester and Hinds \cite{GotayNesterHinds}, which follows Dirac regarding primary first-class constraints (p. 2394). 
 Having developed a sophisticated theory, they rightly turned to applying it to Maxwell's electromagnetism.  Having written the Hamiltonian field equations, they made a transverse-longitudinal split of the 3-vector potential $\vec{A}$ and its canonical momentum.  They obtain, among other familiar results, 
\begin{eqnarray*}
\frac{\partial A_{\perp} }{\partial t} = undetermined, \\
\frac{\partial \vec{A}_L}{\partial t} = -\nabla A_{\perp}.
\end{eqnarray*}
Thus ``the evolution of $A_{\perp}$ and $\vec{A}_L$ is arbitrary.'' \cite{GotayNesterHinds}  So far, so good---at least if one counts a \emph{single} bit of arbitrariness, given that the arbitrariness in $ -\nabla A_{\perp}$ determines the arbitrariness in the evolution of $\vec{A}_L.$  Time will tell if that interpretation is maintained.
\begin{quote}
Let us compare the equations of motion [of which the relevant parts just appeared] and the known gauge freedom of the electromagnetic field with the predictions of the algorithm.\ldots [What the primary constraint generates has as] its effect to generate arbitrary changes in the evolution of $A_{\perp}.$  This is clearly consistent with the field equations.
\end{quote}
 It is consistent with the field equations \emph{if} one pays the price by adding a gradient in  $\frac{\partial \vec{A}_L}{\partial t}$ in accord with the familiar electromagnetic gauge freedom. But that turns out not to be what they have in mind. 
\begin{quote} Turning now to the first-class secondary constraint \ldots, we wonder if it is the generator of physically irrelevant motions.  \ldots [Imposing a suitable \emph{demand}] has the effect of replacing the second of equations [shown above] by 
$$\frac{ d \vec{A}_L}{d t} = - \vec{ \nabla} A_{\perp} - \vec{\nabla} g$$ 
and leaving the others invariant.  As $A_{\perp}$ is arbitrary to begin with, it is evident that this equation is completely equivalent to [the ones shown].  The addition of $-\vec{\nabla} g$ to the right-hand side of this equation has no physical effect whatsoever. \cite[p. 2397]{GotayNesterHinds}. \end{quote}
It is now clear that they envisage two arbitrary functions, not one.  
But this latter physical equivalence claim is  false.  Now that the former claim (about the primary constraint) is disambiguated, it is seen to be false also. 
By taking the divergence of  the modified equation, one sees the falsehood of the second physical equivalence claim:
\begin{eqnarray} 
 \vec{ \nabla} \cdot  \frac{ \partial \vec{A}_L}{\partial t} +     \vec{ \nabla} \cdot   \vec{ \nabla} A_{\perp} +  \vec{ \nabla} \cdot  \vec{\nabla} g = 0 \nonumber \\ 
= \vec{ \nabla} \cdot  \frac{\partial (\vec{A}_L + \vec{A}_T)  }{\partial t} +     \vec{ \nabla} \cdot   \vec{ \nabla} A_{\perp} +  \vec{ \nabla} \cdot  \vec{\nabla} g  \nonumber \\
= \vec{ \nabla} \cdot  \frac{\partial \vec{A}  }{\partial t} +     \vec{ \nabla} \cdot   \vec{ \nabla} A_{\perp} +  \vec{ \nabla} \cdot  \vec{\nabla} g  \nonumber \\
=  \vec{ \nabla} \cdot ( \frac{\partial \vec{A}  }{\partial t} +      \vec{ \nabla} A_{\perp} )+  \vec{ \nabla} \cdot  \vec{\nabla} g \nonumber= \\ 
\vec{ \nabla} \cdot \vec{E}  +  \nabla^2  g =0.
\end{eqnarray}
The authors see their result as a vindication of the extended Hamiltonian formalism for the case of electromagnetism, but actually the electric field is changed by a so-called gauge transformation and Gauss's Law is spoiled. 
This problem illustrates a remark of Henneaux and Teitelboim's:
\begin{quote}
The identification of the gauge orbits with the null surfaces of the induced two-form relies strongly on the postulate made throughout the book that all first-class constraints generate gauge transformations.  If this were not the case, the gauge orbits would be strictly smaller than the null surfaces, and there would be null directions not associated with any gauge transformation. \cite[p. 54]{HenneauxTeitelboim}  
\end{quote}

Another difficulty appears in Faddeev's treatment \cite{FaddeevEnergy}, which, perhaps through notational confusion, gives the impression of showing that the constraint $p^i,_i$ generates a standard electromagnetic gauge transformation.  He uses the symbol $E_k$ for the canonical momentum conjugate to $A_k$. 
(Faddeev does not  introduce a canonical momentum conjugate to $A_0$, so this paragraph will avoid the term ``secondary constraint.'')
 It isn't difficult to show that the canonical momentum $E_k$ has vanishing Poisson bracket with the smeared constraint $\int d^3x \Lambda(x) \partial_k E_k$ for smearing function $\Lambda(x).$  But this result is hardly decisive for \emph{the electric field}.  %
 Taking results about the canonical momentum and treating them as applying to the electric field is, in effect, the fallacy of equivocation regarding the meaning of $E_k$.  Faddeev does not investigate what a Poisson bracket with  $\int d^3x \Lambda(x) \partial_k E_k$   does to $A_0,_i - \dot{A}_i,$ which is what pushes on charges.  Hence the supposed demonstration that $\int d^3x \Lambda(x) \partial_k E_k$ generates an electromagnetic gauge transformation fails.  The relation between the electric field and the canonical momentum in facts holds only \emph{on-shell}, that is, after all Poisson brackets are taken, because it reappears in the equation $\dot{q}=\frac{\delta H}{\delta p}$ after being discarded in the Legendre transformation.  Hence showing that the canonical momentum has vanishing Poisson bracket with $\int d^3 \Lambda(x) \partial_k E_k$
does not show the same result for the electric field. If one hasn't defined  a Poisson bracket for a velocity, one can at least ascertain what the smeared divergence of the canonical 3-momentum does to $A_0,_i$ and $A_i$ and then infer the altered $F_{\mu\nu}$ (as was just done above).  If one defines a Poisson bracket for a velocity (following Anderson and Bergmann \cite{AndersonBergmann}), one can calculate the Poisson bracket of the electric field with the smeared divergence of the canonical 3-momentum and find that it isn't $0$. Thus the smeared divergence of the canonical 3-momentum does not generate a gauge transformation. Instead the smeared divergence of the canonical momentum generates a transformation that \emph{breaks} that very link between that canonical momentum and the electric field, which was supposed to be recovered from $\dot{q}- \delta H/\delta p=0,$  as will be discussed in more detail below.
 But the error seems to be tempting and to pass by without remark.  
 To recover a gauge transformation without $p^0,$ one also needs to impose the change in the Lagrange multipliers $\lambda,$ which in this case means $A_0$  \cite{FradkinVilkoviskyHLEquivalence}. %


\section{Gauge Generator as Tuned Sum of First-Class Constraints}

 While Dirac studies electromagnetism  \cite{DiracLQM}, his process of adding terms to and subtracting terms from the Hamiltonian is not systematic \cite{PonsDirac}.  Neither is there much concern to preserve equivalence to the Lagrangian formalism \cite{DiracCanadian2}.  He seems not to calculate explicitly what his first-class constraints do.  

One can add the two independently smeared constraints' actions together:
\begin{eqnarray} \delta A_{\mu}(x) = \{ A_{\mu}(x), \int d^3y [p^0(y) \xi(t,y)      + p^i,_i(y) \epsilon(t,y)] \}= \delta^0_{\mu} \xi  -\delta_{\mu}^i \partial_i \epsilon, \end{eqnarray}   
getting  their combined change in $\vec{E}$:
\begin{eqnarray} \delta F_{0n } = -\delta \vec{E} =  - \partial_{n}\xi  -   \partial_n \partial_{0} \epsilon.  \end{eqnarray}
If one puts the constraints to work together as a team by setting $\xi=-\dot{\epsilon}$ to make the $\delta F_{0n}=0,$  then   
\begin{eqnarray} \delta A_{\mu}(x) = \{ A_{\mu}(x), \int d^3y [-p^0(y) \dot{\epsilon}(t,y)      + p^i,_i(y) \epsilon(t,y)] \} =  -  \delta^0_{\mu} \dot{\epsilon}  -\delta_{\mu}^i \partial_i \epsilon = -  \partial_{\mu} \epsilon, \end{eqnarray}  which is good. 
  Not surprisingly in light of the gauge generator  \cite{AndersonBergmann,CastellaniGaugeGenerator,PonsSalisburyShepleyYang} \begin{eqnarray} G=\int d^3x (p^i,_i \epsilon - p^0 \dot{\epsilon}),\end{eqnarray} $p^0$ and $p^i,_i$  generate \emph{compensating} changes in $\vec{E}$  when suitably combined.
Indeed we have pieced together $G$  by demanding that the changes in $\vec{E}$ cancel out.  Two wrongs, with opposite signs and time differentiation, make a right.  This tuning, not surprisingly, is a special case of what Sundermeyer found necessary to get first-class transformations to combine suitably to get the familiar gauge transformation for the potentials for Yang-Mills \cite[p. 168]{Sundermeyer}.  Sundermeyer, however, did not calculate the field strength(s) and notice the disastrous spoilage of the Gauss's law-type constraints by first-class transformations.  Hence recovering the familiar gauge transformation of the potentials for him was merely a good idea.

One could make similar remarks about Wipf's treatment of Yang-Mills fields \cite[p. 48]{WipfBadHonnef}, 
 except that Wipf doesn't  seem to find recovering the Lagrangian gauge transformations even a good idea; it's simply an option.  (The same seems to hold for Banerjee, Rothe and Rothe \cite{BanerjeeRotheRothe} \emph{vis-a-vis} \cite{RotheRothe}.) If one doesn't have that taste, one at any rate has ``the canonical symmetries'' from an arbitrary sum of the first-class constraints \cite[p. 48]{WipfBadHonnef}; 
 Wipf advocates extending the Hamiltonian \cite[pp. 40, 41]{WipfBadHonnef} to account for all the gauge freedom. 
But what one actually one gets from an arbitrary sum of first-class constraints is the spoilage of Gauss's law. 

Now with the primary and secondary constraints working together, Gauss's law is preserved:
 $\nabla \cdot \vec{E} =    \nabla^2 \xi  +   \nabla^2 \dot{\epsilon}=   \nabla^2 (-\dot{\epsilon} +\dot{\epsilon})=0.$
 A first-class constraint typically does \emph{not} generate a gauge transformation; it is \emph{part} of the gauge generator $G,$  which  here acts as $\{  A_{\mu},G \}= -\partial_{\mu} \epsilon $,  $\{  p^{\mu},G \} =0.$  

Advocates of the gauge generator $G$ combining the constraints \cite{AndersonBergmann,CastellaniGaugeGenerator,PonsSalisburyShepleyYang} generally have aimed to recover the usual transformation of the \emph{potential(s)} $A_{\mu}$;  the transformation of the field strength(s) $F_{\mu\nu}$ would follow obviously in the usual way and so did not need explicit calculation.  Part of the contribution made here is to calculate the effects of a first-class constraint on the \emph{field strength} $F_{\mu\nu}$, because calculating the effect on the gauge-invariant observable field strength leaves nowhere to hide.  By taking the curl before tuning the sum of first-class constraints rather than after, one sees more vividly why that tuned sum is required and the separate pieces are unacceptable; one sees the looming disaster to be avoided.
The commutative diagram illustrates what differs and what is the same in commuting the operations of inferring $F_{\mu\nu}$ from $A_{\mu}$ and in inferring from effects of the tuned combination $G$ from the effects of the separate first-class constraints:
$$\begin{CD}
A_{\mu}     @>L-equiv.>>  G=\int d^3x(-p^0\dot{\epsilon}+\epsilon p^i,_i) @>>> \delta A_{\mu}= -\partial_{\mu} \epsilon  \\
@V\int d^3x(p^0\xi+\epsilon p^i,_i)VV                      @.                      @VVcurlV \\
 \delta A_{\mu} = \delta^0_{\mu} \xi - \delta^i_{\mu} \epsilon,_i  @>curl>>   \delta F_{\mu\nu}  = (\delta^0_{\nu}  \xi,_{\mu}   - \delta^i_{\nu}  \epsilon,_{i\mu}) - \mu \leftrightarrow \nu   @>L-equiv.>\xi=-\dot{\epsilon}>  \delta F_{\mu\nu}=0
\end{CD}$$
While the top line is fairly familiar, the bottom line appears to be novel, with the merely partial exception of (\cite{CostaDiracConjecture}). It is of course unacceptable to have $\delta F_{\mu\nu} \neq 0,$ so requiring  Lagrangian equivalence from the Hamiltonian resolves the trouble.

This explicit treatment exhibits the force of conditions about the gauge generator that have long been known more abstractly.\footnote{Thanks to Josep Pons for this remark.}  In particular, Hamilton's equations are preserved by a quantity $G$ if and only if $G(t)$ is first-class, $\frac{\partial G}{\partial t} + \{G, H_p\} \equiv pfcc$ (pfcc meaning an arbitrary sum of primary first-class constraints),  and the Poisson bracket of $G$ with the primary first-class constraints is a sum of primary first-class constraints \cite{PonsDirac}.  (Second-class constraints are assumed to be absent or at any rate eliminated.)  
If one attempts to substitute for $G$, for cases like electromagnetism, Yang-Mills, or GR, a primary constraint multiplied by an arbitrary function of time (and space), the equation above yields something of the form $pfcc + sfcc \equiv pfcc$ (sfcc being an arbitrary sum of secondary first-class constraints), which is false.  Likewise, attempting to substitute for electromagnetism (not Yang-Mills or GR, which are more intricate) a secondary constraint multiplied by an arbitrary function gives $sfcc + 0 \equiv pfcc$ \cite[p. 127]{Sundermeyer}, which is also false.  The sum of the two schematic equations,  $(pfcc + sfcc)  + (sfcc +0) \equiv pfcc,$ 
 by contrast, is not obviously hopeless, and indeed works out if one tunes the relative coefficients correctly.


\section{Gauge Invariance of $\dot{q}- \frac{\delta H}{\delta p}= -E_i -p^i=0$  }

The gauge transformation properties of momenta differ between Lagrangian and Hamiltonian formulations. In the Lagrangian formalism, one \emph{defines} the canonical momenta as $ \frac{\partial L}{\partial q^i,_0 }$; they inherit their gauge transformation behavior through this definition.  
In the Hamiltonian formalism, one thing changes and another one doesn't.  What changes is the gauge transformation behavior of $p_i,$ which is independent, so it no longer inherits gauge transformation behavior from  $ \frac{\partial L}{\partial q^i,_0 }. $  Instead $p_i$ gets its gauge transformation behavior somehow or other (together or separately) from Poisson brackets with first-class constraints.  What does not change is the gauge transformation behavior of $\dot{q}^i$, which in many examples is heavily involved in the Lagrangian gauge transformation  behavior of  $\frac{\partial L}{\partial q^i,_0 }$.  

One hopes, of course, to recover from the new Hamilton's equation $\dot{q}^i- \frac{\delta H}{\delta p_i}=0$   what one had in the Lagrangian formalism in $ \frac{\partial L}{\partial q^i,_0 }$ and then gave up in setting the conjugate momenta free.  On the other hand, if one is careless about gauge transformation properties of $p_i$ or (more commonly) $\dot{q}^i$ in the Hamiltonian formalism, it is possible to spoil $\dot{q}^i- \frac{\delta H}{\delta p_i}=0$.  
The equation $\dot{q}^i - \frac{\delta H}{\delta p_i}=0$ holds only on-shell; it is not an identity in the Hamiltonian formalism.  Thus one thing that one must not do (though one sometimes sees it done) is to pretend that one can use this equation to define the gauge transformation properties of $\dot{q}^i$.  One cannot do that, because gauge transformations are generated using Poisson brackets, \emph{i.e.}, off-shell, at the same logical `moment' as the equations 
$\dot{q}^i=\frac{\delta H}{\delta p_i}$, which are also generated using Poisson brackets. Thus there is no relationship between $\dot{q}^i$ and  $\frac{\delta H}{\delta p_i}$ at that stage.  For the case of electromagnetism, there is no relationship between the electric field $\vec{E}$ (which is not quite $\dot{A}_i$, but is close enough) and the canonical momentum $p^i$ (which is not quite $\frac{\delta H}{\delta p^i}$, but, again, is close enough).  On the other hand, one  \emph{still knows}  the gauge transformation behavior of the velocity $\dot{q}^i$, namely, the time derivative of the gauge transformation of $q^i$:  $\delta \dot{q}^i = ( \delta q)^i,_0.
$  For electromagnetism, this means roughly that one can simply calculate how the new $F_{\mu\nu}$ following from the new $A_{\mu}$ by the usual definition (taking the curl), differs from the old $F_{\mu\nu}$ derived from the old $A_{\mu}.$   
The on-shell equality of $\dot{q}^i$ and $\frac{\delta H}{\delta p_i}$ thus imposes a condition of \emph{on-shell equality of the gauge transformations} of $\dot{q}^i$ and $\frac{\delta H}{\delta p_i}$.  This condition restricts what sorts of transformations can be gauge transformations.  In the case at hand, $\vec{E}$ is roughly $\dot{A}_i$ (corrected by some unproblematic spatial derivatives of $A_{\mu}$) and $p^i$ is roughly $\frac{\delta H}{\delta p_i}$ (again, corrected by some unproblematic spatial derivatives of $A_{\mu}$).  Thus the condition is that the gauge-transformation properties of $\vec{E}$ and $p^i$ agree on-shell.  
While $p^i$ has vanishing Poisson bracket with each first-class constraint separately in this case, $\vec{E}$ has vanishing Poisson bracket only with the gauge generator $G$ that combines the two first-class constraints so as to cancel out the change that each one makes separately.  Gauge invariance of $\dot{q}^i=\frac{\delta H}{\delta p_i}$ thus necessitates regarding $G$ as the gauge generator, and not regarding each isolated first-class constraint as generating a gauge transformation.  

For the specific case of electromagnetism, one has the (canonical) Hamiltonian  \cite[p. 127]{Sundermeyer}
\begin{eqnarray} \int d^3x [ \frac{1}{2} (p^i)^2 + \frac{1}{4} F_{ij}^2 - A_0 p^i,_i]. \end{eqnarray}
Thus $\dot{q} -\frac{\delta H}{\delta p}=0$ is just, for three of the four components of $A_{\mu},$ 
\begin{eqnarray}
\dot{A}_i - \frac{\delta H}{\delta p^i} = \dot{A}_i - (p^i + A_0,_i) = \dot{A}_i + A^0,_i - p^i = -E_i -p^i=0.
\end{eqnarray}
What one reckons as gauge freedom must be compatible with this on-shell relationship.  While $p^i$ has vanishing Poisson brackets with each first-class constraint separately, $E_i$ is invariant under a transformation of $A_{\mu}$ only if one tunes the primary and secondary constraints' smearing functions to cancel out the induced changes in $E_i.$  Thus being a gauge transformation requires more than leaving $p^i$ alone (as one might think sufficient if one gives the Hamiltonian formalism priority \cite{CostaDiracConjecture} \cite[p. 20]{HenneauxTeitelboim}); it requires leaving $E_i$ alone as well.  Otherwise one makes the relationship $\dot{A}_i - \frac{\delta H}{\delta p^i}= -E_i -p^i=0$   gauge-dependent, spoiling Hamiltonian-Lagrangian equivalence and undermining the physical meaning of $p^i$ on-shell (the only context where $p^i$ has any physical meaning). 
These concerns about the extended Hamiltonian bear some resemblance to Sugano, Kagraoka and Kimura's \cite{SuganoExtended}. 
Likewise, Banerjee, Rothe and Rothe connect restrictions on the gauge parameters, the expected Lagrangian gauge transformations, preserving the Hamilton equations of motion, and the gauge generator \cite{BanerjeeRotheRothe}.   
Pons also  derives conditions for the gauge generator $G$ by requiring the gauge-covariance of Hamilton's equations \cite{PonsDirac}.  
The force of such conditions is made more evident by providing an explicit example in which such conditions are often unwittingly violated.


\section{First-Class Constraints and Invariance of the Canonical Action}

Within the context of Lagrangian field theory it is well known that a gauge transformation  is the sort of transformation that changes the action either not at all or at most by a boundary term, thus preserving the equations of motion (at least by the old standards in which one did not worry much about boundary terms).  
When one has introduced canonical momenta $p^i$ conjugate to $A_i$ as auxiliary fields, one can use the canonical action $\int dt d^3x  (p \dot{q}- \mathcal{H})$ as a slightly unusual Lagrangian,  the canonical Lagrangian, varied so as to get Hamilton's equations as Euler-Lagrange equations.  One thereby preserves manifest Hamiltonian-Lagrangian equivalence.  One can also introduce a dummy canonical momentum $p^0$ conjugate to $A_0$, the variation of which gives the vacuous equation of motion $\dot{A}_0 = \dot{A}_0$; thus one accommodates the primary constraint $p^0$ and the big Poisson bracket that involves $A_0.$ What becomes of the criterion of changing the action by at most a boundary term in the Dirac-Bergmann context?

The primary Hamiltonian augments the canonical Hamiltonian density $\mathcal{H}_c$ with the term $p^0 \dot{A}_0$; the velocity in effect keeps its job of describing the time variation of $A_0$ because the new momentum $p^0$ is constrained to vanish \cite[pp. 92-94]{SudarshanMukunda}.  The canonical action is 
\begin{eqnarray}
S_H = \int dt d^3x (p^{\mu} \dot{A}_{\mu} - \mathcal{H}_p )  = \nonumber  \\
 \int dt d^3x (p^{i} \dot{A}_{i} - \mathcal{H}_c )  = \nonumber  \\
\int  dt d^3x (p^{i} \dot{A}_{i} - \frac{1}{2}p^{i2} -p^i A_0,_i - \frac{1}{4} F_{ij}^2).
\end{eqnarray}

One can calculate with Poisson bracket of the canonical Lagrangian  with the  smeared primary first-class constraint  $\int d^3y p^0(y) \xi(t,y).$  (Apart from a spatial boundary term that traditionally one discarded without guilt \cite[p. 127]{Sundermeyer}, the spatial  integration $\int  d^3x$ in the canonical Lagrangian does nothing, so the calculation could be carried out with the canonical Lagrangian density instead.  Still less does one need temporal integration of the Lagrangian to get the action.)
One has
\begin{eqnarray} 
 \{  \int d^3y p^0(y) \xi(t,y), \int  d^3x [ p^i(x) \dot{A}_i(x) - \mathcal{H}_c(x)] \} = \nonumber  \\
   \int d^3y  d^3x \xi(t,y) \{  p^0(y),   - \mathcal{H}_c(x) \} = \nonumber  \\
\int d^3y  d^3x \xi(t,y) \{ \frac{1}{2} p^{i2}(x) + p^i A_0,_i(x) + \frac{1}{4} F_{ij}^2(x) , p^0(y)   \} = \nonumber  
\int d^3y  d^3x  \xi(t,y) \{  p^i A_0,_i(x), p^0(y)   \} = \nonumber  \\
\int d^3y  d^3x [- \xi(t,y) p^i,_i(x) \{   A_0, p^0(y)   \}] = \nonumber  \\
\int d^3y  d^3x [- \xi(t,y) p^i,_i(x) \delta^0_0 \delta(x,y)]  = \int   d^3x [- \xi(t,x) p^i,_i]. 
  \end{eqnarray} 
  This is, of course, neither $0$ identically nor a boundary term.  It does vanish using the secondary first-class constraint $p^i,_i=0$ \cite{LusannaNoether}.  But to preserve equivalence to the Lagrangian formalism, one wants strong, not weak invariance (or quasi-invariance).  Hence the primary first-class constraint does not generate a Lagrangian-equivalent gauge transformation.
This argument has the virtue of making no assumption about the meaning or set of examples of observables.

What does the secondary first-class constraint do to the canonical action? One might prefer to see what it does to the Lagrangian density or the Lagrangian, but temporal integration provides a resource for evading the Poisson bracket of a velocity.  
 One has  
\begin{eqnarray} 
\int dt \{  \int d^3y \epsilon(t,y) p^i,_i(y) , \int  d^3x [ p^j(x) \dot{A}_j(x) - \mathcal{H}_c(x)] \} = \nonumber  \\
\int dt d^3y d^3x [ \{  \epsilon(t,y) p^i,_i(y) ,  p^j(x) \dot{A}_j(x) \}
 -   \{  \epsilon(t,y) p^i,_i(y) ,   \mathcal{H}_c(x) \} ] = \nonumber  \\ 
\int dt d^3y d^3x [ \{  \epsilon(t,y) p^i,_i(y) ,   p^j(x) \dot{A}_j(x) \}
 -   \{  \epsilon(t,y) p^i,_i(y) ,   \frac{1}{2}p^{j2}(x) + p^j A_0,_j(x) + \frac{1}{4} F_{jk}^2(x) \} ] = \nonumber  \\ 
\int dt d^3y d^3x [ \{  \epsilon(t,y) p^i,_i(y) ,   p^j(x) \dot{A}_j(x) \}
 - \frac{1}{2}  \{  \epsilon(t,y) p^i,_i(y) ,  F_{jk}(x)   \} F_{jk}(x)] = \nonumber  \\ 
\int dt d^3y d^3x [ \{  \epsilon(t,y) p^i,_i(y) ,  p^j(x) \dot{A}_j(x) \}
 - \frac{1}{2}  \epsilon(t,y),_i \{   p^i(y) ,  A_{k}(x)   \} F_{jk},_j(x)] = \nonumber  \\ 
\int dt d^3y d^3x [ \{  \epsilon(t,y) p^i,_i(y) ,  p^j(x) \dot{A}_j(x) \}
+ \frac{1}{2} \epsilon(t,y),_i  \delta^i_k \delta(x,y)   F_{jk},_j(x) ] = \nonumber  \\ 
\int dt d^3y d^3x  \{  \epsilon(t,y) p^i,_i(y) ,  p^j(x) \dot{A}_j(x) \}
- \frac{1}{2} \int   dt d^3x \epsilon(t,x)    F_{jk},_{jk}(x) = \nonumber  \\ 
 - \int dt d^3y d^3x  \{  \epsilon(t,y),_i p^i(y) ,  p^j(x) \dot{A}_j(x) \}
- 0. 
\end{eqnarray}

The first term has been carried along because a little more thought is useful in evaluating it.  One approach is to integrate by parts with respect to time, avoiding the mysterious Poisson bracket with  $\dot{A}_i(x)$ by changing the action by a boundary term, then calculating the Poisson bracket of $p^i(y)$ with $A_j(x),$ then integrating by parts back to pull the time differentiation off $\dot{p}^j(x).$  
The result of this effort to avoid taking a Poisson bracket of a velocity, which works  for the canonical action but not for the canonical Lagrangian density, is, after a spatial integration by parts as well,
$$-\int dt d^3x p^i,_i(x) \dot{\epsilon}(t,x).$$  This is, of course, not $0$ identically or even a boundary term; the secondary first-class constraint, again, does not generate a gauge transformation.  It does, however, cancel the result from the primary constraint above if one chooses $\xi = -\dot{\epsilon}$---that is, if one puts the two constraints to work together as the gauge generator $G$.
Unlike the electric field argument, this action-based  argument type works for other theories also, including theories for which the concept of observables is contested, such as General Relativity \cite{GRChangeNoKilling}.

%
%


\subsection{Counting Degrees of Freedom}

One might think that correct counting of degrees of freedom would depend on whether one takes the generator of gauge transformations to be a special combination of the first-class constraints or an arbitrary combination.  In the former case, there are only as many independent functions of time (and perhaps space) as there are \emph{primary} first-class constraints; some of the constraints are smeared with the time derivative of functions that smear other constraints.  In the latter case there are as many independent functions of time (and perhaps space) as there are first-class constraints.  However, behavior over time is irrelevant; hence a function and its time derivative, being independent \emph{at a moment}, count separately.  Thus the counting works out the same either way \cite[pp. 89, 90]{HenneauxTeitelboim}.  Getting the correct number of degrees of freedom thus does not show whether each first-class constraint or only the special combination $G$ generates gauge transformations.

 %
\section{Neglect of Secondaries in Hamiltonian}

One major reason that first-class constraints wrongly have been thought to generate gauge transformations is that Dirac claims to prove it early in his book \cite[p. 21]{DiracLQM}.  One finds the same proof repeated in other works  \cite{HenneauxTeitelboim,WipfBadHonnef,RotheRothe}.
The canonical Hamiltonian is, up to a boundary term \cite[p. 127]{Sundermeyer},
\begin{eqnarray} \int d^3x \left( \frac{1}{2} p^{i2} + \frac{1}{4} F_{ij}^2 - A_0 p^i,_i \right). \end{eqnarray}
The primary Hamiltonian adds the primary constraint with an arbitrary velocity $v$ or $\dot{A}_0$. Dirac, not using the gauge generator $G$,  saw the arbitrary velocities $v$ multiplying the primaries outside his $H^{\prime}$ but apparently forgot  the corresponding arbitrary $q$'s (like $A_0$) multiplying the secondaries \emph{inside} $H^{\prime}$.  
Thus he did not notice that the first-class primaries  outside $H^{\prime}$ and first-class secondaries inside $H^{\prime}$ work as a team to generate gauge transformations.
If one considers the time evolution of $A_j$, one has
\begin{eqnarray}
\dot{A}_j = \{A_j,H_p \} = \{ A_j, H_c \} = p^j + A_0,_j;
\end{eqnarray}
the second, gauge-dependent term comes from the secondary constraint in $H_c.$
  Resuming with Dirac, 
\begin{quote}  [w]e come to the conclusion that the $\phi_a$'s, which appeared in the theory in the first place as the primary first-class constraints, have this meaning:  \emph{as generating functions of infinitesimal contact transformations, they lead to changes in the $q$'s and the $p$'s that do not affect the physical state.} \cite[p. 21, emphasis in the original]{DiracLQM} \end{quote}
One could hardly reach such a conclusion without thinking that the primaries were the locus of all dependence on the arbitrary functions. But in fact the secondary constraint plays a role with that same arbitrary function $\dot{A}_0$ (integrated over time).
Dirac then conjectures that the same gauge-generating property that he ascribes to primary first-class constraints also holds for first-class secondaries.  As appeared above, neither the primaries nor the secondaries  generate a gauge transformation in electromagnetism. (The simple direct test made above has the advantage of already knowing what a gauge transformation in electromagnetism is, whereas Dirac is confronting the harder problem of figuring out in the general case what a gauge transformation is.)  
  Dirac's error presumably encouraged him to extend the Hamiltonian in order to recover what was apparently missing \cite[pp. 25, 31]{DiracLQM}.  
 But extending the Hamiltonian is unnecessary (because the secondary constraints and full gauge freedom are already there) and obscures the relation of the fields to those in the more perspicuous and reliable Lagrangian formalism \cite[p. 39]{IsenbergNester100}. Indeed the extended Hamiltonian breaks Hamiltonian-Lagrangian equivalence \cite{GraciaPons}. 
Requiring Hamiltonian-Lagrangian equivalence fixes the supposed ambiguity permitting the extended Hamiltonian \cite{SuganoEquivalence}.

Pons's reworking of Dirac's analysis of gauge transformations diagnoses and avoids  Dirac's mistake \cite{PonsDirac}.  Pons, like Dirac, takes the two gauge-related trajectories to have identical initial conditions---not merely physically equivalent ones related by a gauge transformation at the initial moment.  As a result, their analyses as applied to electromagnetism would make the $A_0$ the same on the two trajectories at the initial moment---thereby making the contribution from the secondary constraint disappear initially because its relative coefficient is $0$.  One can make this assumption at the initial moment, but one cannot impose it (without serious loss of generality) a second time.  Pons leaves room for the secondary constraints within $H^{\prime}$ to play a role because integrating $v^0=\dot{A}_0$ will make the values of $A_0$ differ between the two trajectories later on.  
Dirac, alas, oversimplifies by forgetting that setting the very same initial data between the two cases implies assuming gauge-dependent entities such as $A_0$ in electromagnetism and the lapse and shift vector in GR to be initially equal (not merely gauge-equivalent).  Dirac's second transformation thus omits the role of the secondaries in $H^{\prime}$ at a time when, unlike the initial moment, one may no longer assume the values of $A_0$ (the secondaries' coefficients) on the two evolutions to be equal  without loss of relevant generality.  
While a more direct analysis (one that does not subtract two evolutions and then set the coefficients of the secondary constraint equal, annihilating the secondary's influence) works at first infinitesimal order, as shown above, Dirac's argument cancels crucial terms at first order in $\delta t$.  Thus the argument works correctly only when run to second order  \cite{PonsDirac}, and the role of the secondary-primary team, not the primaries alone, is evident.


\subsection{Perpetuation  in Recent Works}

Dirac's argument is widely followed \cite{Govaerts,HenneauxTeitelboim,WipfBadHonnef,RotheRothe}.  The problem will be clear if one starts with Wipf's treatment; those by Govaerts \cite[pp 116, 117]{Govaerts} and Rothe and Rothe \cite[p. 68]{RotheRothe} are basically the same, while Henneaux and Teitelboim's is a bit too brief for complete clarity in isolation.  The time evolution of a system with first-class constraints is derived from the primary Hamiltonian $H_p$ (the canonical Hamiltonian $H$ plus the primary constraints $\phi_a$ with arbitrary multiplier functions $\mu^a$).  For a phase space quantity $F,$ Wipf says that one compares 
\begin{quote}  
two infinitesimal time evolutions of $F=F(0)$ given by $H_p$ with different values of the multipliers, 
\begin{eqnarray*}
 F_i(t) = F(0) + t\{F,H\}
 + t\{F, \phi_a\} \mu^a_i \hspace{.3in} i=1,2  \hspace{.3in}.  \hspace{1in}  (5.16)
\end{eqnarray*}
The difference $\delta F=F_2(t)-F_1(t)$ between the values is then 
\begin{eqnarray*}
\delta_{\mu}F = \{F, \mu^a \phi_a\}, \hspace{.3in},  \hspace{.3in}  \mu=t(\mu^2 - \mu^1).  \hspace{1in} (5.17)
\end{eqnarray*}
Such a transformation does not alter the  physical state at time $t$, and hence is called a [\emph{sic}] \emph{infinitesimal gauge transformation}. [reference to Dirac's book \cite{DiracLQM} in arxiv version, which has other slight differences also] \cite[p. 40]{WipfBadHonnef} 
\end{quote}
Like Dirac, Wipf has overlooked the fact that the canonical Hamiltonian also is influenced by the multiplier functions:  the canonical Hamiltonian contains the gauge-dependent quantity $A_0$ multiplying the secondary constraint, while the multiplier function is $\dot{A}_0.$  Thus not only the $\mu^a$ multiplier functions, \emph{but also the canonical Hamiltonian} $H$, needs a subscript $1$ or $2$---at least after the initial moment when one can stipulate away that difference by assuming \emph{identical} (not merely equivalent) initial data.  With this mistake corrected, one has in general
 \begin{eqnarray}
\delta_{\mu}F = t\{F, H_2-H_1\} +   t \{F,  \phi_a\} (\mu^a_2 - \mu^a_1) = \nonumber \\
 t \{F, \int d^3y  (-A^2_0+A^1_0)(y) \pi^i,_i(y) \}      + t \{F,  \int d^3y p^0(y) \} (\mu_2 - \mu_1). 
\end{eqnarray}
The correct expression exhibits the secondary constraint(s) working together with the primary constraint(s). One can cancel out the term $ t \{F, \int d^3y  (-A^2_0+A^1_0)(y) \pi^i,_i(y) \} $ only in special cases, such as
at the initial moment.  Given the restricted erroneous expression involving only the primary constraint, a `gauge transformation' that changes only $A_0$ would be exhibited. But as was shown in detail above, or as follows from electrostatics, changing $A_0$ while leaving everything else alone \emph{does} alter the physical state, and hence is not a gauge transformation.   It is obvious that this expression does not change the canonical momenta $p^0$ or $p^i$; what does it do to $A_{\nu}$?  The corrected expression, unlike Dirac's, changes $A_j$ as well, as it should, and affects the initial data also. 
Letting $F=A_{\nu}(x)$ gives (changing notation from $t$ to $\delta t$ for a small interval, and recalling that our initial moment can be called $t=0$)
\begin{eqnarray}
\delta_{\mu} A_{\nu}(\delta t, x) = \nonumber \\ \delta t \{A_{\nu}(0,x), \int d^3y  (-A^2_0+A^1_0)(0,y) \pi^i,_i \}      + \delta t \{A_{\nu},  \int d^3y p^0 \} (\mu_2 - \mu_1) = \nonumber \\
\delta t \int d^3y \delta^i_{\nu} \delta(x,y)(A_0^2,_i - A_0^1,_i)(y) + \delta t \delta^0_{\nu}(\mu_2-\mu_1)(x) = \nonumber \\
\delta t  \delta^i_{\nu} (A_0^2,_i - A_0^1,_i)(x) + \delta t \delta^0_{\nu}(\dot{A}_0^2 -\dot{A_0^1})(0,x) = \nonumber \\
\delta t   (A_0^2 - A_0^1),_{\nu}(0,x).
\end{eqnarray}
This expression clearly resembles the usual gauge transformation property of electromagnetism $- \partial_{\nu} \epsilon$, so one can say that the two evolutions differ by a (standard) gauge transformation, as one would hope.
Thus it is false that the primary first-class constraints generate a gauge transformation in examples like electromagnetism, because it is a special combination of the primaries and secondaries that does so. 
The primary by itself changes $\vec{E},$ as does the secondary by itself.
Continuing with Wipf,
\begin{quote}
[w]e conclude that the most general physically permissible motion should allow for an arbitrary gauge transformation to be performed during the time evolution.  But $H_p$ contains only the primary FCC.  We thus have to add to $H_p$ the secondary FCC multiplied by arbitrary functions.  This led Dirac to introduce the \emph{extended Hamiltonian}\ldots which contains \emph{all} FCC [reference to Dirac's book \cite{DiracLQM}]. $H_e$ accounts for all the gauge freedom.

Clearly, $H_p$ and $H_e$ should imply the same time evolution for the classical observables.
\cite[pp. 40, 41]{WipfBadHonnef}  
\end{quote}  
But the secondary first-class constraint already is present in the primary Hamiltonian, as is the gauge freedom, so there is nothing missing that needs adding in by hand.  
Such facts are all the more consequential in relation to General Relativity, in which the canonical Hamiltonian is nothing but secondary constraints (and boundary terms).

Now the problem in the treatment of Henneaux and Teitelboim can be identified readily and treated briefly.  
\begin{quote}
Now, the coefficients $v^a$ are arbitrary functions of time, which means that the value of the canonical variables at $t_2$ will depend on the choice of the $v^a$ in the interval $t_1 \leq t \leq t_2.$ Consider, in particular, $t_1 + \delta t.$  The difference between the values of a dynamical variable $F$ at time $t_2$, corresponding to two different choices $v^a,$ $\tilde{v}^a$ of the arbitrary functions at time $t_1,$ takes the form
\begin{eqnarray*}
\delta F = \delta v^a [F,\phi_a]  \hspace{1in} (1.35)
\end{eqnarray*}
with $\delta v^a = (v^a - \tilde{v}^a) \delta t.$ Therefore the transformation (1.35) does not alter the physical state at time $t_2.$ We then say, extending a terminology used in the theory of gauge fields, that \emph{the first-class primary constraints generate gauge transformations}. \cite[p. 17]{HenneauxTeitelboim}
\end{quote}
By now the problem is clear:  the secondary constraints also appear in the Hamiltonian, with coefficients involving the same arbitrary function integrated over time, $-A_0$ for the case of electromagnetism.  At the first instant this term's effects are cancelled by the assumption of identical initial data, but beyond lowest infinitesimal order that cancellation fails and the role of the primary-secondary team with related coefficients involving the same arbitrary function $\dot{A}_0$ appears.  
%

%

\section{Dirac Conjecture's Presupposition}
Dirac, having supposedly shown that primary first-class constraints generate gauge transformations, conjectured that secondary first-class constraints do the same \cite{DiracLQM}.  Eventually it was found that this conjecture has counterexamples, namely ineffective constraints, though they are a bit exotic and might sensibly be banned \cite{HenneauxTeitelboim}. 
But the Dirac conjecture has a much more serious problem, namely, the falsehood of its presupposition that primary first-class constraints generate gauge transformations.  Whether that problem makes the Dirac conjecture false or lacking in truth value will depend on the logical details of the formulation, but it certainly winds up not being an interesting truth. Complementing the falsification by direct calculation above is a diagnosis (just above) of the mistake that Dirac and others have made in failing to pay attention to the term $\int d^3x (- A_0 p^i,_i)$ term in the Hamiltonian.

How does one reconcile this result that a primary first-class constraint does not generate a gauge transformation with the multiple `proofs' of the Dirac conjecture in the literature \cite{GitmanTyutin,HenneauxTeitelboim,CaboDiracConjecture,SobolevDiracConjecture,RotheDirac} and the statements that it can be made true by interpretive choice \cite{GotayIneffective,HenneauxTeitelboim}?  
These proofs are of more than one type.  Some presuppose that a Dirac-style argument has already successfully addressed primary first-class constraints, so the only remaining task involves secondary or higher order constraints.  The remaining task tends to involve statements about first-class constraints, which are simply \emph{assumed} to generate gauge transformations individually.  Thus `proofs' of the Dirac conjecture are frequently just statements about Poisson brackets and first-class secondary (and higher) constraints---straightforward technical questions with results that are, presumably, correct, though the failure of primary first-class constraints typically to generate gauge transformations blocks the inference that secondary first-class constraints generate them as well---which is what Dirac conjectured \cite[p. 24]{DiracLQM}.  

Other proofs  draw attention to the fact that the Dirac conjecture, at least as it is now treated, is 2-sided, and hence can fail in two different ways.  One way for the Dirac conjecture, as now treated, to fail would be for a first-class constraint \emph{not to play any part} in generating a gauge transformation.  That is an issue in counterexamples involving ineffective constraints \cite{RotheDirac}.  Hence proofs that all first-class constraints do in fact play a role in generating gauge transformations under suitable conditions are indeed of interest for confirming \emph{part} of the Dirac conjecture as it is now discussed---though whether Dirac himself would have been disappointed to learn that some first-class constraints don't generate anything at all is unclear, given that doing nothing at all leaves the world unchanged, as a gauge transformation should.  When those proofs involve teaming up the first-class constraints such that there are only as many arbitrary smearing functions as \emph{primary} first-class constraints, rather than as many as there are first-class constraints ($1$ \emph{vs.} $2$ in electromagnetism, $4$ \emph{vs.} $8$ in GR, \emph{etc.}), then those proofs, while of interest in exploring the connection between first-class constraints and gauge transformations, nonetheless  refute what Dirac actually conjectured, as in (\cite{RotheDirac,GitmanTyutinEquivalent}).


\section{Primordial Observable  $E^i$ \emph{vs.} Auxiliary  $p_i$}

While it is acknowledged that the extended Hamiltonian is not equivalent to $L$ strictly, this inequivalence is often held to be harmless because they are equivalent for ``observables.'' This claim presumably is intended to mean that the extended Hamiltonian is \emph{empirically equivalent} to $L,$ differing only about unobservable matters.  
Such a response will be satisfactory only if ``observable'' here is used in the ordinary sense of running experiments.  Technical stipulations about the word ``observable,'' especially distinctively Hamiltonian stipulations, are irrelevant.  Unfortunately it is not the case that the extended Hamiltonian is empirically equivalent to the Lagrangian, a fact that has been masked by equivocating on the word ``observable'' between the ordinary experimental sense and a technical Hamiltonian sense.   
It is peculiar to think of observing canonical momenta conjugate to standard Lagrangian coordinates---in fact it seems to be impossible to observe that kind of canonical momentum as such. What would be the operational procedure for observing $p^i$?   Rather, its experimental significance is purely on-shell, parasitic upon the observability of suitable functions of $q^i$ and/or  derivatives of $q^i$---derivatives (spatial and temporal) of $A_{\mu}$ in the electromagnetic case. 
One neither acquires new experimental powers (such as the ability to sense canonical momenta) nor loses old ones (such as the ability to detect a certain combination of derivatives of $A_{\mu}$) by changing formalisms from the Lagrangian to the Hamiltonian.  There are two ways to see that $p^i$ is not the primordial observable electric field. The first way involves the fact that $p^i$ does not even appear as an independent field in the Lagrangian formalism, which formalism is correct and transparent.
  While it is perfectly acceptable for some quantity to be introduced that is on-shell equivalent to the Lagrangian electric field, there is no way for that new quantity to become the electric field primordially, rather than merely derivatively and on-shell.  $A_{\mu}$ or a function of its derivatives still has that job. Apart from constraints, canonical momenta are auxiliary fields in the Hamiltonian action $\int dt(p\dot{q}-H(q,p))$ \cite{HenneauxAuxiliary}:  one can vary with respect to $p$, get an equation $\dot{q}-\frac{\delta H}{\delta p}=0$ to solve for $p$, and then eliminate $p$ to get $\int dt L$.
One would scarcely call an auxiliary field a primordial observable and the remaining $q$ in $L$ derived. 
The second way involves the fact that the electric field is what pushes on charge; but it is easy to see that in both the Lagrangian and Hamiltonian contexts, what couples to the current density is not $p^i$, but $A_{\mu}$.  For a complex scalar field $\psi$,   the Lagrangian interaction term takes the form $\sim (\psi\partial_{\alpha}\psi^* - \psi^*\partial_{\alpha}\psi)A^{\alpha} + \psi \psi^* A^2.$ The absence of terms connecting $\psi$ with derivatives of $A_{\mu}$ implies that charge couples to $A_{\mu}$ and/or its derivatives, not to the canonical momenta conjugate to $A_{\mu},$ even in the Hamiltonian context.  
 What is the operational procedure for measuring $p^i$?  The only plausible answer is to use on-shell equivalence to the empirically available $F_{0i}$, which involves derivatives of $A_{\mu}$.  Otherwise, what reason is there to believe that any procedure for measuring $p^i$ involves a measurement of the quantity that pushes on charge?  Thus one \emph{should} be disturbed, \emph{pace} Costa \emph{et al.} \cite{CostaDiracConjecture}, by the failure of $\dot{A}_i = \frac{\delta H_E}{\delta p_i}$ to return the usual Lagrangian relation between $p^i$ and the derivatives of $A_{\mu}$ from the extended Hamiltonian.  
 The coupling of charge-current to $A_{\mu}$ ensures that $A_{\mu}$ or something built from its derivatives is the primordial observable electric field.  
Thus the usual argument \cite{CostaDiracConjecture,HenneauxTeitelboim,WipfBadHonnef,RotheRothe}  to show that the inequivalence of the extended Hamiltonian to the Lagrangian is harmless because irrelevant to observable quantities, fails. Unless ``observables'' are taken in the ordinary empirical sense, rather than a technical Hamiltonian sense, empirical equivalence is not shown.

The proof of the Dirac conjecture by Costa \emph{et al.} \cite{CostaDiracConjecture} deserves special comment. This paper goes beyond other treatments of the supposed equivalence of the extended Hamiltonian to the primary Hamiltonian for observables \cite{HenneauxTeitelboim,RotheRothe} in explicitly addressing the example of electromagnetism in sufficient detail.  
The equivalence conclusion is reached by  explicitly taking the canonical momentum $p^i$ to be the primordial physically meaningful quantity playing the role of the electric field.  For a function of canonical coordinates and momenta (no time derivatives), having vanishing Poisson bracket with the gauge generator requires having vanishing Poisson bracket with each first-class constraint, because different orders of time derivative of the smearing function cannot cancel each other out \cite{CostaDiracConjecture}.  
But that latter condition opens the door to taking all first-class constraints to generate gauge transformations and using the extended Hamiltonian, they claim.  They recognize that one can use Hamiltonian's equations from the primary Hamiltonian and find a quantity that is equal in value on-shell to a gauge-invariant function of $q$ and $p$.  I observe that the electric field is in this category.
They also observe that such a quantity is invariant under the gauge generator of the primary Hamiltonian (the specially tuned combination of first-class constraints) and is not invariant under the first-class constraints separately: 
\begin{quote} 
[o]ne can verify the invariance under [the usual electromagnetic gauge transformation of $A_{\mu}$] of the equations of motion \ldots
\begin{eqnarray*} \partial^0 A^j = \pi_j + \partial^j A^0,  \hspace{1in} (3.8b) 
\end{eqnarray*}
\ldots deriving from the primary Hamiltonian\ldots.

We next recognize $F^{ij},$ $\pi_j$ \ldots [matter terms suppressed] as the canonical forms of the basic gauge-invariant quantities of electrodynamics.  One can easily check that all these functions are indeed first class.  Thus, $F^{ij},$ $\pi_j$\ldots are also invariant under the extended infinitesimal transformations [generated by an arbitrary sum of independently smeared first-class constraints].
\ldots [That extended first-class transformation] leaves invariant the equations of motion\ldots
\begin{eqnarray*}
\partial^0 A^j = \pi_j + \partial^j A^0 -\partial^j\xi^2,  \hspace{1in} (3.12b) 
\end{eqnarray*}
\ldots arising from the extended Hamiltonian
\begin{eqnarray*}
H_E = H + \int d^3x\{\xi^1({\bf x})\pi_0({\bf x}) + \xi^2({\bf x})[\partial^j\pi_j({\bf x}) -\ldots ]\}. \hspace{1in} (3.13)
\end{eqnarray*} [spinor contribution in secondary constraint suppressed] \\
Here $\xi^1$ and $\xi^2$ are arbitrary Lagrange multipliers.

As a matter of fact, the sets of equations of motion (3.8) and (3.12) are \emph{different}.  However, irrespective of whether one starts from (3.8) or (3.12) one arrives at the Maxwell equations
\begin{eqnarray*}
\partial^0 F^{ij} = \partial^i \pi_j - \partial^j\pi_i, \hspace{1in}  (3.14)   \\
\partial^0 \pi_j = \partial^i F^{ij} \ldots, \hspace{1in} (3.15) 
\end{eqnarray*} 
 \cite[pp. 407, 408]{CostaDiracConjecture} \end{quote} 
I note  the absence of Gauss's law!

They continue:
\begin{quote}
Therefore, $H_T$ and $H_E$ generate the same time evolution for the gauge-invariant quantities, as required by [the equation of motion for gauge invariant phase space functions].

We now discuss the alternative formalism-dependent realizations of the electric field ($-\pi_j$).  From (3.8b) one obtains
\begin{eqnarray*}
\pi_j = F^{0j}. \hspace{1in} (3.17)
\end{eqnarray*}
Hence, $F^{0j}$ is a faithful realization of $\pi_j$ within the formalism of the primary Hamiltonian.  We can check that $F^{0j}$ is invariant under [the gauge generator related to the primary Hamiltonian, which combines the first-class constraints with related smearings] but not under [the sum of separately smeared first-class constraints, which is related to the extended Hamiltonian formalism].
\cite[p. 408]{CostaDiracConjecture} \end{quote}
\emph{This is the crucial point announced in my paper's title}---but Costa \emph{et al.} fail to recognize the absurdity of the results of the extended Hamiltonian formalism.  They continue:
\begin{quote}
On the other hand, the formalism of the extended Hamiltonian provides the equally faithful realization for $\pi_j$ [see Eq. (3.12b)] 
\begin{eqnarray*}
\pi_j = F^{0j} + \partial^j\xi^2, \hspace{1in}  (3.18)
\end{eqnarray*}
which is invariant under [the sum of independently smeared first-class constraints].  One should not be puzzled by the fact that (3.18) does not coincide with (3.17) or, what amounts to the same thing, with the Lagrangian definition of $\pi_j$ \ldots. \cite[p. 408]{CostaDiracConjecture}  
\end{quote}
But one \emph{should} be puzzled, because their identification of primordial \emph{vs.} formalism-dependent observable quantities is exactly backward. 
 If $\pi_j$ is equated to the electric field (as they say), and if $F^{0j}$ is just an abbreviation for a familiar expression involving derivatives of $A_{\mu}$ (as follows from (3.12b) and (3.18)---and hence is still the electric field, I note!),  then we have the contradiction 
(electric field = electric field + arbitrary gradient). With this contradiction in hand, one can derive various other plausible errors. This arbitrary gradient  is what spoiled Gauss's law above.  
In any case  $F^{0j}$ has a much better claim to be the electric field than does $\pi_j,$ which is just an auxiliary field in the Hamiltonian action.  Thinking that functions of phase space were the only quantities that needed to stay gauge invariant---that is, not considering the actual electric field---is what opened the door to the extended Hamiltonian and taking each first-class constraint as separately generating a gauge transformation.  One should infer that an isolated first-class constraint does not generate a gauge transformation in electromagnetism.  
 $F^{0j}$  is the primordial observable electric field; the canonical momentum as an independent field is formalism-dependent, not even appearing  in the Lagrangian formalism.  In a Lagrangian for charged matter with an electromagnetic field, charge-current couples primordially to $A_{\mu}$, from which $\vec{E}$ is derived, and not to the canonical momentum. Velocities (such as appear in the electric field) are not physically recondite---automobiles have gauges that measure them---but canonical momenta are: they acquire physical significance solely on-shell, as Costa \emph{et al.} remind us.   Hence failure to recognize the fundamentality of the Lagrangian formalism leads them to claim to have vindicated the Dirac conjecture, when they had all the ingredients and calculations necessary to refute it instead.

Crucial to gauge-transforming the electric field (as opposed to the canonical momentum to which it is equal on-shell) is having a gauge transformation formula for velocities.  In a Hamiltonian formalism it is tempting, though inadvisable, to avoid velocities in favor of functions of $q$ and $p$.  But the Lagrangian formalism essentially involves the commutativity of gauge variation and time differentiation \cite{BanerjeeRothe,BanerjeeRotheRothe}.  Imposing that condition in the Hamiltonian formalism using the primary Hamiltonian (the one equivalent to the Lagrangian) yields the gauge generator $G$ \cite{BanerjeeRothe,BanerjeeRotheRothe}.  Thus the Hamiltonian formalism naturally can give the correct gauge transformation for velocities and quantities built from them, such as the electric field.  One does not need to avoid looking for gauge-invariant quantities involving the velocities and default to functions of only $q$ and $p$ in a Hamiltonian context, as Costa \emph{et al.} did \cite{CostaDiracConjecture}.  Alternately, one can be satisfied in a (primary) Hamiltonian formalism with functions of $q$ and $p$ \cite{ChaichianPhysicalQuantities} but, in view of the need to preserve Hamiltonian-Lagrangian equivalence, avoid seeking the largest collection of transformations (the first-class transformations rather than just the gauge generator $G$) that preserve the phase space quantities at the expense of Hamiltonian-Lagrangian equivalence.


\section{Canonical Transformations Generating Field Redefinitions}

  None of this confusion associated with Hamiltonian transformations that aren't induced by Lagrangian gauge transformations should be much of a surprise, ideally, in that Anderson and Bergmann explicitly discussed how the preservation of the Lagrangian constraint surface, which they called $\Sigma_l,$ corresponds to canonical transformations generated by the gauge generator $G$ \cite{AndersonBergmann}.  Hence one would expect transformations that aren't generated by $G$---\emph{e.g.}, those generated by an isolated primary constraint in a theory (such as Maxwell's electromagnetism or GR) where the gauge generator $G$ doesn't contain that primary constraint in isolation (\emph{i.e.}, smeared by its very own arbitrary function)---not to preserve the Lagrangian constraint surface.  Hence the point that a first class constraint by itself (in theories where such does not appear in isolation in $G$) generates not a gauge transformation, but a violation of the usual Lagrangian constraint surface, is already implicit in  Anderson and Bergmann---at least if one is working with canonical transformations.  (Outside the realm of canonical transformations, one can still take Poisson brackets directly.  But then there are far fewer rules and hence there is much less reason to expect anything good to happen.)   
As they observe,
\begin{quote}
Naturally, other forms of the hamiltonian [\emph{sic}] density can be obtained by canonical transformations; but the arguments appearing in such new expressions will no longer have the significance of the original field variables $y_A$ and the momentum densities defined by Eq. (4.2) [which defines the canonical momenta as $\pi^A \equiv \frac{ \partial L}{\partial \dot{y}_A }$].  It follows in particular that transformations of the form (2.4) [``invariant'' transformations changing $\mathcal{L}$ by at most a divergence, such as electromagnetic gauge transformations or passive coordinate transformations in GR] will change the expression (4.9) [for the Hamiltonian density] at most by adding to it further linear combinations  of the primary constrains, i.e., by leading to new arbitrary functions $w^i.$  \cite[p. 1021]{AndersonBergmann} \end{quote}
 So they invented the gauge generator $G$ to make sure that the $q$'s and $p$'s keep their usual meanings.

Let us recall Dirac's words to motivate seeking a connection to canonical transformations:
\begin{quote}
We come to the conclusion that the \ldots primary first-class constraints, have this meaning:  \emph{as generating functions of infinitesimal contact transformations, they lead to changes in the $q$'s and the $p$'s that do not affect the physical state}.  \cite[p. 21]{DiracLQM}
\end{quote} 
 Above I have found that first-class constraints generate, by direct taking of Poisson bracket, a bad physical change.  Yet contact transformations (presumably the same as canonical transformations \cite{Goldstein}) are well known to be permissible changes of description for the same physics.
That is a feature of dynamics in general, not Dirac-Bergmann constrained dynamics in particular.
  A plausible resolution of this tension between the physical equivalence expected from the phrase ``contact transformations'' and the bad physical changes  
is that Dirac is not actually making canonical transformations.  It will be useful, then, to use smeared a first-class constraint as a generating function in electromagnetism  to see what results, how it relates to Dirac's treatment, and how it compares to gauge transformations.  
Such a use employs $p^0,$ but not the fact that $p^0=0$ (its being a constraint) or its having vanishing Poisson brackets with the other constraints and Hamiltonian (its being first-class).  Physical equivalence is preserved, but  only by losing some of the original fields' meanings by making an awkward position-dependent field redefinition, it will turn out.

A physically equivalent, canonically transformed action results from requiring that Hamilton's principle be satisfied for both $\int dt (p \dot{q} - H)$ and for $\int dt (P \dot{Q} -K)$ and requiring the two integrands to differ by only a total derivative \cite[p. 380]{Goldstein}.  Let $C= \int d^3 y \epsilon(t,y) p^0(y).$  One can add to the Hamiltonian action the time integral of the total time derivative of this quantity.  One gets new canonical coordinates, $Q^A = q^A + \frac{\delta C}{\delta p_A},$ and new canonical momenta, $P_A = p_A - \frac{\delta C}{\delta q^A},$ and a slightly altered Hamiltonian, $K= H+ \frac{\partial C}{\partial t}=H + \int d^3y p_0 \frac{\partial \epsilon }{\partial t},$ which adds a term proportional to a primary constraint only.  Of the new $Q$'s, only the 0th differs from the old $q$'s ($ Q^0 = q^0 +  \epsilon)$;  the new momenta are the same as the old.  The trouble arises subtly:  for the other $Q$'s velocity-momentum relation,  $\dot{Q^a} = \frac{ \delta K}{\delta P_a},$ the dependence on the 0th canonical coordinate in $K$ involves the altered $Q^0. $  
  The electromagnetic scalar potential is involved in the relation between $\dot{A}_i$ and $p^i,$ so changing the scalar potential alters the relationship between the canonical momenta and the velocities, the sort of issue to which Anderson and Bergmann called attention.    For $q^0$ corresponding to $A_0$ (or the lapse $N$ or shift vector $N^i$ in General Relativity), one can change $q^0$ \emph{alone} however one likes over time and place (which is what the corresponding primary constraint does)---but only at the cost of ceasing to interpret the new canonical coordinate $Q^0=q^0 + \delta q^0$ as (minus\footnote{I use $-+++$ metric signature.  Indices are placed up and down freely, depending on whether the general paradigm $Q^A$ or the specific case $A_{\mu}$ is more relevant. }) the scalar potential $A_0$ (or lapse $N$ or shift $N^i$)!
The new Hamiltonian $K$ differs from $H$ only by a term involving a primary constraint $p_0=P_0,$ which doesn't matter.  The new velocity-momentum relationship is 
\begin{eqnarray}
\dot{Q}^i = \frac{ \delta K}{\delta P_i } = \frac{ \partial}{\partial P_i} (\frac{1}{2} P_j^2 + \frac{1}{4}F_{jk}^2 + P_j \partial_j[Q^0-\epsilon])   =   P_i + \partial_i(Q^0 -\epsilon). 
 \end{eqnarray}
One can solve for $P_i$ and then take the 3-divergence:
\begin{eqnarray}  P_i,_i = \partial_i(\dot{Q}^i -Q^0,_i + \epsilon,_i) = \partial_i(\dot{q}^i -\partial_i q^0) = \partial_i F_{0i} = -\partial_i E_i. \end{eqnarray} 
\emph{By using the full apparatus of a canonical transformation and keeping track of the fact that $Q^0$ is no longer (up to a sign) the electromagnetic scalar potential} as $q^0$ is, one can resolve the contradiction about vanishing \emph{vs.} nonvanishing divergence of the canonical momentum \emph{vis-a-vis} the electric field.  Such reinterpretation, which strips the new canonical coordinates of some of their usual physical meaning and replaces them with a pointlessly indirect substitute, though mathematically permitted, \emph{is certainly not what people usually intend when they say that a first-class constraint generates a gauge transformation}.  What they mean, at least tacitly, is that the fields after the transformation by direct application of Poisson brackets (not a canonical transformation) have their usual meaning---hence one would (try to) calculate the electric field from $\dot{Q}^i -Q^0,_i$ (thus spoiling the Lagrangian constraints, as shown above) rather than $\dot{Q}^i -Q^0,_i + \epsilon,_i.$
  Supposing that one attempts to retain the old connection between the 0th canonical coordinate and the electromagnetic scalar potential, one can calculate the alteration in the electric  field (that is, the electric field from $Q^A$ less the electric field from $q^A$) as
$ \delta  F_{0n} = \partial_0 \delta A_n - \partial_n \delta A_0  = 0 - \partial_n \frac{ \delta C}{\delta p_0 }= -\partial_n \epsilon,$ as found above by more mundane means.  To avoid the contradiction of a physics-preserving transformation that changes the physics, one can and must re-work the connection between $Q^0$ and $A_0,$ as shown.  But simply avoiding this sort of generating function, one that is not (a special case of) $G$, is more advisable.

 In short, as a canonical transformation generating function with suitable smearing, $p_0$, the primary first-class constraint, generates only an \emph{obfuscating position-dependent change of variables}.  It has nothing to do with the usual gauge freedoms of electromagnetism (or GR, by analogy).  It has nothing to do with $p_0$'s being first-class; the canonical transformation would work equally well for Proca's massive electromagnetism, in which that constraint is second-class.  Only in detail does it even depend on $p_0$'s being a constraint, as opposed to merely something that lives on phase space.  It is easy to see reasons not to make such transformations, and wrong to make them without understanding what they do.  
   

One can also try the (smeared) secondary constraint $p^i,_i$ as a generator of a canonical transformation:
$D= \int d^3y [-\epsilon,_i p^i(y)]$ after dropping a boundary term.   
The new canonical coordinates are $Q^A = q^A + \frac{\delta D}{\delta p_A} = q^A - \epsilon,_i \delta^i_A = A_{\alpha} -  \epsilon,_i \delta^i_{\alpha}.$   The new  canonical momenta are  $P_A = p_A - \frac{\delta D}{\delta q^A}= p_A.$  One sees that the new $Q^i$ are not the original electromagnetic 3-vector potential $A_i$ anymore.  (They are not a gauge-transformed vector potential, either, unless one throws the trouble onto $Q^0$ by stripping it of its relation to the electromagnetic scalar potential.)  
The new Hamiltonian is  $K= H+ \frac{\partial D}{\partial t}=H + \int d^3y [ - p^i \epsilon,_{0i}],$ which differs from the old by a term proportional to the secondary constraints (and perhaps a boundary term).   
Thus the altered $\dot{Q}-P$ relation is $\dot{Q}^i = \frac{\delta K}{\delta P_i} = P_i + Q^0,_i -\epsilon,_{0i}.$  One can take the divergence and solve for $P^i,_i:$  
$P^i,_i = \partial_i(Q^i,_0 - Q^0,_i + \epsilon,_{0i})  =  \partial_i(q^i,_0 - \partial_i q^0) = \partial_i F_{0i} = - \partial_i E_i.$
By  taking into account the fact that the new $Q$'s are no longer all just the electromagnetic 4-vector potential $A_{\mu}$,  one resolves the contradiction between vanishing and nonvanishing divergence.  The electric field $\vec{E}$, which is an observable by any reasonable standard, is no longer specified simply by (derivatives) of the new canonical coordinates $Q$, but requires the arbitrary smearing function $\epsilon$ used in making the  change of field variables also. That is permissible but hardly illuminating.

One can do basically the same thing with Proca's massive electromagnetism \cite{Sundermeyer,GitmanTyutin} with mass term $- \frac{m^2}{2} A_{\mu} A^{\mu}$, taking the secondary constraint, now second-class, as the generator of a canonical transformation.  The secondary sprouts a new piece $m^2 A_0$. The transformed massive Hamiltonian $K$ gets an extra new term   $ m^2 Q^0 \dot{\epsilon}.$ The new canonical momenta reflect a change in the primary constraint form: $P_0 = p_0 -m^2 \epsilon.$  But everything cancels out eventually, leaving equations equivalent to the usual ones for massive electromagnetism, naturally.   Only in detail does the first-class (massless) \emph{vs.} second-class (massive) character of the secondary constraint make any difference.  As the generator of a canonical transformation, a first-class constraint doesn't generate a gauge transformation in massless electromagnetism any more than a second-class constraint generates a gauge transformation in massive electromagnetism.   Both generate  permissible but pointless field redefinitions. 

 The key difference is that a special combination of first-class constraints in massless electromagnetism does generate a gauge transformation, whereas in massive electromagnetism, there is no gauge transformation to generate, so no combination of anything can generate one.  Amusingly, given that the key issue is changing $A_{\mu}$ by a four-dimensional gradient, and not directly the first-class or even constraint character of the generator, one can use      
the same special sum  $\int d^3y [-p^0(y) \dot{\epsilon}(t,y)      + p^i,_i(y) \epsilon(t,y)] $  as applied to massive electromagnetism to generate a Stueckelberg-like gauged version of massive electromagnetism, with the smearing function $\epsilon$, in this case not varied in the action, as the gauge compensation field.  $\int d^3y [-p^0(y) \dot{\epsilon}(t,y)      + p^i,_i(y) \epsilon(t,y)] $  is no longer a sum of constraints (not even second-class ones, though $p^0$ is a second-class constraint).  This possibility might take on some importance in application to installing artificial gauge freedom in massive Yang-Mills theories, where the proper form has been a matter of some controversy \cite{Ruegg,YMembed,UmezawaKamefuchi,Salam,GrosseKnetter,BanerjeeGhosh}.

Finally, one can use the gauge generator $G$ as the generator of a canonical transformation in Maxwell's electromagnetism.  It turns out that, in contrast to an arbitrary function on phase space (or a first-class constraint) as a generator, the gauge generator $G$ generates the \emph{very same thing} for the canonical variables as a canonical transformation as it does `by hand' by taking the Poisson bracket directly with $q$ and $p$.  Dropping a spatial divergence, one has $G=\int d^3x [-\epsilon,_{\mu} p^{\mu}].$  One gets the new canonical coordinates $Q^A = q^A + \frac{\delta G}{\delta p_A}  = A_{\alpha} - \epsilon,_{\alpha} $ and new canonical momenta $P_A = p_A - \frac{\delta G}{\delta q^A}=p_A,$ and a slightly altered Hamiltonian, $K= H+ \frac{\partial G}{\partial t}=H + \int d^3y   [ - p^{\mu} \epsilon,_{\mu0}],$ which adds related terms  proportional to the primary and secondary constraints (and a spatial boundary term).  %
 Significantly, $Q^A - q^A = \frac{\delta G}{\delta p_A} = \{q^A, G\}$ and $P_A - p_A = - \frac{\delta G}{\delta q^A} = \{ p_A, G\}.$ That is, $G$ does the very same thing to $q^A$ and $p_A$ whether one simply takes the Poisson bracket with $G$ directly or uses $G$ to generate a canonical transformation. Thus if one uses $G$, one can be nonchalant (as Dirac was about separate first-class constraints) about whether one is making a canonical transformation or is merely directly taking a Poisson bracket; that lack of concern should not carry over to expressions different from $G$, however.    $G$ does one good thing, recovering the usual electromagnetic gauge transformations,  used either way.  By contrast,   each isolated  first-class constraint offers a choice of two bad things (one disastrous, one merely awkward):  it can either destroy the field equations if used directly in Poisson brackets, or generate a confusing change of physical meaning of the variables as the generator of a canonical transformation.

One can summarize in a table some of the results about using the gauge generator $G$ \emph{vs.} a smeared individual constraint or other phase space function,  and using it  as a canonical transformation generating function \emph{vs.} using it directly \emph{via} Poisson bracket.  Presumably the experience for electromagnetism largely carries over for other constrained theories.  For the first-class theory one has these phenomena: 
\begin{center} 
\begin{tabular} {| c | c | c |} \hline
    & Canonical transformation   & Direct Poisson bracket  \\ \hline
Gauge generator $G$  & Gauge transformation  & Gauge transformation   \\ \hline
Smeared  constraint & Locally varying  field redefinition  &  Spoils  $\vec{\nabla} \cdot \vec{E} = 0$ \\ \hline
\end{tabular} 
\end{center}
The entries in the first column can be described in more detail.  
 One can illustrate the illuminating (invariant) canonical transformations related to $G$ (top left corner) and the obscuring but permissible more general canonical transformations (bottom left corner) in the following  diagrams.

 The first is a commutative diagram with well understood entries and transformations.  
(The equation numbers correspond to the remarks in Anderson and Bergmann \cite{AndersonBergmann}.) 
\vspace{.25in}
$$\begin{CD}
\mathcal{L}    @>invariant\hspace{.05in}gauge\hspace{.05in}2.4:>\delta\mathcal{L}=div,\hspace{.05in}\delta{A}_{\mu}=\partial_{\mu}\xi,\hspace{.05in}{\delta}g_{\mu\nu}=\pounds_{\xi}g_{\mu\nu}> \mathcal{L}^{\prime}  \\
@Vconstrained\hspace{.05in}LegendreVV @VVconstrained\hspace{.05in}LegendreV \\
\mathcal{H}  @>invariant\hspace{.05in}canonical\hspace{.05in}G>preserves\hspace{.05in}q_A\hspace{.05in}sense,\hspace{.05in}4.2:\hspace{.05in}\pi^A=\frac{\partial\mathcal{L}}{\partial\dot{q}_A}> \mathcal{H}^{\prime}
\end{CD}$$

One can of course also make point transformations, changes among the $q_A$'s only. In electromagnetism, one might use $A^{\mu}$ instead of $A_{\mu}$; that is probably the least bad choice if one does not stick with $A_{\mu}.$  In GR one is free to use $g_{\mu\nu}$, $\mathfrak{g}^{\mu\nu}$ (which equals $ g^{\mu\nu} \sqrt{-g}$), or various other fields, for example.  For Anderson and Bergmann, this freedom to make point transformations is already implied by their rather abstract use of $q_A$ (or actually $y_A$ in their notation) and rather general form of gauge transformations.  A field redefinition from one choice of $q_A$ to another will of course induce a contragredient change in the canonical momenta. 
One can also add a divergence to the Lagrangian density.  Such an alteration will also tend to alter the canonical momenta, but not mysteriously.  These two changes were combined to simplify the primary constraints of GR in 1958 \cite{AndersonPrimary,DiracHamGR}.  One could augment the diagram above to indicate more fully the resources of Lagrangian field theory.  The main point, however, is to distinguish adequately what is allowed within the Lagrangian formalism from the greater, and more dangerous, generality of the Hamiltonian formalism.

The second is an unhealthy aspiring commutative diagram illustrating how allowing general canonical transformations---for example, a single primary or secondary first-class constraint---leads to entries and transformations that are not widely understood, if meaningful at all.  \vspace{.25in}
$$\begin{CD}
\mathcal{L}    @>?>> \mathcal{L}^{\prime}  \\
@Vconstrained\hspace{.05in}LegendreVV @AAinverse\hspace{.05in}constrained\hspace{.05in}Legendre?A \\
\mathcal{H}  @>general\hspace{.05in}canonical\hspace{.05in}>violates\hspace{.05in}q_A\hspace{.05in}sense\hspace{.05in}or\hspace{.05in}4.2:\hspace{.05in}\pi^A=\frac{\partial\mathcal{L}}{\partial\dot{q}_A}> \mathcal{H}^{\prime}
\end{CD}$$
A canonical transformation to action-angle variables, for example, would give a Hamiltonian that would prohibit an inverse Legendre transformation back to a Lagrangian \cite[p. 80]{SudarshanMukunda}.  
Suffice it to say that Hamiltonian-Lagrangian equivalence is obscured by general canonical transformations.  It is not very obvious what the resulting equations mean physically, given that the usual Lagrangian variables such as $A_{\mu}$ and $g_{\mu\nu},$ not the canonical momenta, are the ones with known direct empirical meaning.  General canonical transformations are useful tricks in mechanics, where one already understands what everything means, but needs to solve specific problems.  But a position-dependent change of variables when one is already on marshy ground, having difficulty identifying change or observables, is inadvisable without the greatest care.

%

\section{Conclusion}

Carefully doing Hamiltonian calculations for electromagnetism, as an end in itself, would be using a sledgehammer to crack a peanut.  But the pattern of ensuring that the Hamiltonian formalism matches the Lagrangian one, which is perspicuous and correct, will prove very illuminating for the analogous treatment of GR.  There the right answers are generally not evident by inspection, and the calculations are difficult and error-prone.  Knowing what a properly dotted ``i'' and a properly crossed ``t'' look like will be crucial in GR, where various attractive entrenched errors related to the first-class-constraint-generates-a-gauge-transformation theme need to be diagnosed.  
In particular, one should use the primary Hamiltonian and its associated gauge generator $G,$ not the extended Hamiltonian and each first-class constraint smeared separately.  
There might be some clarification achieved for canonical quantization.



%

\section{Acknowledgments}

Thanks to Jeremy Butterfield, George Ellis, Claus Kiefer,  Josep Pons, Don Salisbury, Kurt Sundermeyer, and David Sloan for comments and encouragement.

%
%
\end{document}